\documentclass[aps,preprint,showpacs,groupedaddress,floatfix]{revtex4}

\usepackage{graphicx}
\usepackage{dcolumn}
\usepackage{bm}

\begin{document}

\title{Pairing transitions in finite-temperature relativistic Hartree-Bogoliubov theory}
\author{Y. F. Niu 
$^1$}
\author{Z. M. Niu 
$^2$}
\author{N. Paar$^{3}$}
\author{D. Vretenar$^3$}
\author{G. H. Wang 
$^1$}
\author{J. S. Bai 
$^1$}
\author{J. Meng 
$^{4,5,6}$}
\email{mengj@pku.edu.cn}

\affiliation{$^1$ Institute of Fluid Physics, China Academy of Engineering
             Physics, Mianyang 621900, China}
\affiliation{$^2$School of Physics and Material Science, Anhui University,
             Hefei 230039, China}
\affiliation{$^3$Physics
Department, Faculty of Science, University of Zagreb, Croatia}
\affiliation{$^4$State Key Laboratory of
Nuclear Physics and Technology, School of Physics, Peking
University, Beijing 100871, China}
\affiliation{$^5$School of Physics and Nuclear Energy Engineering,
Beihang University, Beijing 100191, China}
\affiliation{$^6$Department of Physics, University of Stellenbosch, Stellenbosch 7602, South Africa}

\date{\today}
\begin{abstract}

We formulate the finite-temperature relativistic Hartree-Bogoliubov theory for spherical nuclei based on a point-coupling functional, with the Gogny or separable pairing force. Using the functional PC-PK1, the framework is
applied to the study of pairing transitions in Ca, Ni, Sn, and Pb isotopic chains.
The separable pairing force reproduces the gaps calculated with the Gogny force not only at zero temperature, but also at finite temperatures. By performing a systematic calculation of the even-even Ca, Ni, Sn, and Pb isotopes, it is found that the critical temperature for a pairing transition generally follows the rule $T_c =0.6 \Delta_n(0)$, where $\Delta_n(0)$ is the neutron pairing gap at zero temperature. This rule is further
verified by adjusting the pairing gap at zero temperature with a strength parameter.

\end{abstract}
\pacs{21.10.-k, 21.60.Jz, 27.40.+z, 27.60.+j, 27.70.+q, 27.80.+w } \maketitle
\date{today}

\section{Introduction}

At finite temperature metal superconductors undergo a phase transition~\cite{Bardeen1957}.
While the notion of a phase transition is well defined for infinite systems, finite many-body
systems also exhibit a phase-transitional behavior although surface effects and statistical fluctuations
tend to smooth out the transition~\cite{Langanke2005,Mulken2001}. For instance, in a warm nucleus
superfluidity vanishes when temperature increases. This is
easily understood in terms of the shell model~\cite{Egido2000}. By increasing temperature nucleons
are excited from levels below the Fermi surface to levels above, resulting in level blocking, and hence
pairing correlations disappear. Experimental evidence has been found in the S-shaped curve of heat capacity as a function of temperature, obtained from level density at low angular momenta~\cite{Melby1999,Schiller2001,Melby2001,Guttormsen2003}. Furthermore, the critical temperature for the quenching of pair correlations is found at $T_c \simeq 0.5 $ MeV for $^{161,162}$Dy, $^{171,172}$Yb~\cite{Schiller2001}, and $^{166,167}$Er~\cite{Melby2001}. In finite-temperature mean-field theory, the vanishing of pairing correlations with increasing temperature occurs as a sharp phase transition at the critical temperature. The critical temperature is calculated to be $T_c = 0.57 \Delta (0)$ in the finite-temperature BCS theory with a constant pairing force $G$~\cite{Sano1963}, and $T_c = 0.5 \Delta (0)$ using a simplified degenerate model~\cite{Goodman1981}, where $\Delta(0)$ is the pairing energy gap at zero temperature.
The effects of statistical fluctuations have been studied in the spirit of the Landau theory~\cite{Levit1984,Goodman1984,Dang1993}, the static path approximation~\cite{Alhassid1984,Rossignoli1994,Rossignoli1996}, as well as the shell model Monte Carlo method~\cite{Lang1993,Dean1995,Langanke1996,Liu2001,Kaneko2004,Langanke2005}. Although large fluctuations
appear for the nuclear system, clear signatures of the pairing transition can still be found even if the sharp phase transition obtained in the mean field approach is smoothed out~\cite{Dean1995,Langanke1996,Liu2001,Langanke2005,Bozzolo1985}. It has been shown that the transition temperature calculated in the shell model is in good agreement with predictions for the critical temperature obtained in the BCS approximation~\cite{Kaneko2004}.

The disappearance of superfluidity with temperature in nuclei was first studied using the BCS theory, and the  pairing transition was predicted at the critical temperature $T_c = 0.57 \Delta (0)$ for  the case of a constant pairing force~\cite{Sano1963}. Later the finite-temperature Hartree-Fock-Bogoliubov (FTHFB) equations were derived~\cite{Goodman1981,Egido1993}, and their BCS limit was obtained. The finite-temperature BCS (FTBCS) equations were solved for a degenerate model, demonstrating that a transition from a superfluid state to a normal state occurs with increasing temperature, and the critical temperature was calculated: $T_c = 0.5 \Delta (0)$~\cite{Goodman1981}. Consequently, the FTHFB model with a pairing-plus-quadrupole Hamiltonian was applied to the study of shape and pairing transitions in rare-earth nuclei, and it was found that the critical temperature for the pairing transition is in the interval $0.5-0.6 \Delta(0)$~\cite{Goodman1986}.
More recently, the BCS or Bogoliubov calculations with self-consistent mean fields have been employed to study  pairing transitions in hot nuclei. In the framework of nonrelativistic theories, the finite-temperature Skyrme Hartree-Fock with the BCS pairing model was employed to investigate the nuclear shell gaps at finite temperatures, of interest for the astrophysical $r$ process~\cite{Reis1999}. The FTHFB method based on the finite-range density-dependent Gogny force, which yields both the particle-hole and particle-particle matrix elements, was applied in the analysis of the behavior of nuclear shell effects, such as pairing correlations and shape deformations, with excitation energy~\cite{Egido2000,Martin2003}. FTHFB calculations with zero-range forces were also performed in studies of pairing correlations in hot nuclei~\cite{Khan2004,Khan2007}, using the mean field  obtained from a Skyrme force and a density-dependent zero-range pairing interaction. Only few studies based on covariant density functionals have been reported so far. The finite-temperature relativistic Hartree-BCS theory with nonlinear interactions has been applied to a study of the temperature dependence of nuclear shapes and pairing gaps for $^{166}$Er and $^{170}$Er~\cite{Agrawal2000}. The temperature was also included in the Dirac Hartree-Bogoliubov theory using the Matsubara formalism~\cite{Lisboa2007,Lisboa2010}.

Because for nuclei far from stability the BCS approximation presents only a poor approximation, the relativistic Hartree-Bogoliubov (RHB) model has extensively been used in studies of nuclei far from $\beta$ stability, including exotic systems with extreme isospin values~\cite{Kucharek1991,Meng1996,Gonzalez-Llarena1996,Meng1998NPA,Serra2002,Niksic2002, Vretenar2005,Meng2006PPNP,Niksic2011,Zhou2010,Li2012,Li2012CPL,Chen2012}.
A number of interesting structure phenomena have been investigated such as the neutron halo in light and medium-heavy nulcei~\cite{Meng1996,Poschl1997,Meng1998,Meng2002PRC}, ground-state properties of deformed proton emitters~\cite{vretenar1999}, the reduction of the effective spin-orbit interaction in drip-line nuclei~\cite{Lalazissis1998}, shape coexistence phenomena in neutron-deficient nuclei~\cite{Lalazissis1999}, restoration of pseudospin symmetry in exotic nuclei~\cite{Meng1998PRC,Meng1999}, new magic numbers in superheavy nuclei~\cite{Zhang2005}, and the occurrence of the halo phenomenon in deformed nuclei~\cite{Zhou2010}.

Details of calculated nuclear properties depend on the choice of the effective relativistic mean-field (RMF) Lagrangian in the particle-hole (p-h) channel, and the treatment of pairing correlations. In recent years relativistic functionals have been developed that are based on zero-range point-coupling interactions ~\cite{Nikolaus1992,Burvenich2002,Niksic2008}, in which the traditional meson-exchange RMF effective interactions are replaced by local four-point (contact) interactions between nucleons. The RMF point-coupling models produce results comparable to those obtained in the meson-exchange representation~\cite{Niksic2011}. Very recently a new nonlinear point-coupling effective interaction has been introduced~\cite{Zhao2010} (PC-PK1) that successfully describes properties of infinite nuclear matter and finite nuclei, including ground states and low-lying excited states~\cite{Zhao2010,Mei2012,Hua2012,Niu2013}. In particular, the empirical isospin dependence of binding energies along either the isotopic or the isotonic chains is reproduced by PC-PK1, making it suitable for applications in exotic nuclei. This parametrization of the relativistic Lagrangian will be our choice for the $p-h$ channel in the present study. For the particle-particle ($p-p$) channel the Gogny force is very successful in the description of pairing correlations~\cite{Decharge1980,Berger1984,Berger1991}. The results are often used as a benchmark for more microscopic investigations~\cite{Kucharek1989,Serra2001}. No cutoff parameter in momentum space is necessary for this pairing force because of its finite range. Recently, a considerably simpler pairing force, separable in momentum space, was introduced~\cite{Tian2009}. It is carefully adjusted to the nuclear matter pairing gap calculated with the Gogny force. The new pairing force is rather simple so that matrix elements in finite nuclei can be expressed as a finite sum of separable terms, while at the same time the cutoff problem of other separable or zero-range pairing forces is avoided. In the present paper we employ these two forces as the pairing interaction in the $p-p$ channel. The validity of the new separable force will be tested at finite temperature in a comparison with the Gogny force.

In the framework of BCS theory, for both superconductivity of metals~\cite{Bardeen1957} and superfluidity of atomic nuclei~\cite{Sano1963}, a linear relationship between the critical temperature $T_c$ and the pairing gap at zero temperature $\Delta(0)$ can be derived with the assumptions of a constant pairing force $G$ in some energy interval around the Fermi surface, as well as a constant single-particle level density $g$ with $gG \ll 1$. The resulting critical temperature is $T_c = 0.57 \Delta(0)$, determined by setting the finite-temperature pairing gap to zero. In the BCS theory only particles in time-reversed orbitals can form a Cooper pair, whereas the more general Bogoliubov theory incorporates additional correlations and thus two particles from different single-particle orbitals can also form a pair. This could lead to a higher critical temperature at which all the correlated pair states are broken. In addition, because of the shell structure of single-particle states the level density is not a constant, so deviations from a linear relation between the critical temperature and the zero-temperature pairing gap can also be induced. It will be, therefore, interesting to investigate  in the Bogoliubov theory the relation between the critical temperature and the pairing gap at zero temperature.

In this work the finite-temperature RHB (FTRHB) framework for spherical nuclei, based on point-coupling functionals with the Gogny or separable pairing force, will be formulated. The newly developed approach is used to study  pairing transitions in Ca, Ni, Sn and Pb isotopes, using the effective interaction PC-PK1. In Sec.~\ref{sec2} the formalism for the finite-temperature point-coupling RHB model is briefly outlined. In Sec.~\ref{sec3} the thermal properties of Sn isotopes, as well as the systematic behavior of the critical temperature for Ca, Ni, Sn and Pb isotopes, are computed and discussed. Finally, Sec.~\ref{sec4} contains a summary and a brief outlook.

\section{Theoretical framework}
\label{sec2}

The minimization of the grand canonical potential yields the finite-temperature (FT) HFB equation. For the details of the derivation we refer the reader to Refs.~\cite{Goodman1981,Egido1993}. The FTRHB equation in the quasiparticle basis reads:
 \begin{equation}
 \label{FTRHB}
  \left( \begin{array}{cc} h_{ll'} -\lambda - M &
 \Delta_{ll'} \\ -\Delta_{ll'}^* & -h_{ll'}^* + \lambda + M
 \end{array} \right)
 \left( \begin{array}{c} U_{l' k} \\ V_{l'k} \end{array} \right) =
 E_k \left( \begin{array}{c} U_{lk} \\ V_{lk} \end{array} \right) \;.
 \end{equation}
 When nucleons are described as Dirac fermions,
$h$ denotes the single-nucleon Dirac Hamiltonian, and $\Delta$ is the pairing field which sums up particle-particle correlations. $M$ is the nucleon mass, and the chemical potential $\lambda$ is determined by the particle number subsidiary condition, i.e., the expectation value of the particle number operator equals the number of nucleons. The column vectors denote the quasiparticle spinors, and $E_k$ are the quasiparticle energies. The single-nucleon Dirac Hamiltonian takes the form \cite{Nikolaus1992,Burvenich2002,Niksic2008,Zhao2010}
 \begin{equation}
 h = \bm{\alpha}\cdot \bm{p} + V + \beta (M+S),
 \end{equation}
 where the local scalar potential $S$ and the time component of vector potential $V$ read
 \begin{eqnarray}
   \begin{array}{l}
     S=\Sigma_{S}+{\tau}_3 \Sigma_{TS3},\\
     V =\Sigma_{V}^0+ \tau_3 \Sigma_{TV3}^0 \;,
   \end{array}
\end{eqnarray}
respectively. The isoscalar-scalar $\Sigma_{S}$, the third component of isovector-scalar $\Sigma_{TS3}$, the time component of isoscalar-vector $\Sigma_{V}^0$, as well as the third and time component of isovector-vector $\Sigma_{TV3}^0$ self-energies are defined by the following relations:
\begin{eqnarray}
   \begin{array}{l}
     \Sigma_{S}=\alpha_S \rho_S+\beta_S\rho_S^2+\gamma_S\rho_S^3+\delta_S\Delta\rho_S, \\
     \Sigma_{TS3}=\alpha_{TS}\rho_{TS3}+\delta_{TS}(\Delta \rho_{TS3}),\\
     \Sigma_{V}^0=\alpha_V \rho_V+\gamma_V \rho_V^3+\delta_V(\Delta \rho_V)+eA^0~\frac{1-\tau_3}{2},\\
     \Sigma_{TV3}^0=\alpha_{TV}\rho_{TV3}+\delta_{TV}(\Delta \rho_{TV3}).\\
   \end{array}
\end{eqnarray}
$\alpha$, $\beta$, $\gamma$, and $\delta$ in the various spin-isospin channels denote the coupling constants (adjustable parameters) that determine a given effective interaction, such as PC-PK1. $A^0$ is the Coulomb field. The single-nucleon densities in the FTRHB theory are computed using the relations
\begin{eqnarray}
  \rho_{S}(\bm{r}) &=& \sum_{E_k>0} V_k^\dagger(\bm{r}) \gamma^0 (1-f_k) V_k(\bm{r}) + U^T_k(\bm{r}) \gamma^0
  f_k U^*_k(\bm{r}), \\
  \rho_{TS3}(\bm{r}) &=& \sum_{E_k>0} V_k^\dagger(\bm{r}) \gamma^0 \tau_3(1-f_k) V_k(\bm{r}) + U^T_k(\bm{r})
  \gamma^0 \tau_3
  f_k U^*_k (\bm{r}),\\
  \rho_{V} (\bm{r})&=& \sum_{E_k>0} V_k^\dagger (\bm{r})(1-f_k) V_k(\bm{r}) + U^T_k(\bm{r})
  f_k U^*_k(\bm{r}), \\
  \rho_{TV3}(\bm{r}) &=& \sum_{E_k>0} V_k^\dagger (\bm{r}) \tau_3(1-f_k) V_k (\bm{r})+ U^T_k(\bm{r})
 \tau_3
  f_k U^*_k(\bm{r}), \\
  \rho_{c} (\bm{r})&=& \sum_{E_k>0} V_k^\dagger(\bm{r}) \frac{1-\tau_3}{2}(1-f_k) V_k(\bm{r}) + U^T_k(\bm{r})
 \frac{1-\tau_3}{2}
  f_k U^*_k(\bm{r}).
 \end{eqnarray}
The thermal occupation probability of quasiparticle states is defined by
 \begin{equation}
 f_k = \langle \alpha_k^\dagger \alpha_k \rangle =  \frac{1}{1+e^{\beta
 E_k}},
 \end{equation}
where $E_k$ is the quasiparticle energy and $\beta = 1/k_{\rm B}T$. $k_{\rm B}$ is the Boltzmann constant and $T$ is the temperature.

The pairing potential reads
 \begin{equation}
 \Delta_{ll'} = \frac{1}{2} \sum_{kk'} V^{pp}_{ll'kk'} \kappa_{kk'},
 \end{equation}
with the pairing tensor at FT,
 \begin{equation}
\kappa = V^*(1-f) U^T + UfV^\dagger.
 \end{equation}
For the pairing interaction we employ two kinds of forces, namely the Gogny force and the separable force. The pairing part of the Gogny force has the form~\cite{Berger1991},
  \begin{equation}
 V^{pp}(1,2) = \sum_{i=1,2} e^{-[(\bm{r}_1-\bm{r}_2)/\mu_i]^2} (W_i + B_iP^\sigma - H_i P^\tau -M_i P^\sigma
 P^\tau),
 \end{equation}
with the set D1S~\cite{Berger1991} for the parameters $\mu_i$, $W_i$, $B_i$, $H_i$, and $M_i$ $(i=1,2)$.
The separable force is from Ref.~\cite{Tian2009},
\begin{equation}
  V(\bm{r}_1,\bm{r}_2,\bm{r}'_1,\bm{r}'_2) = -G
  \delta(\bm{R}-\bm{R}') P(r) P(r') \frac{1}{2} (1-P^\sigma),
 \end{equation}
where $\bm{R}=\frac{1}{2}(\bm{r}_1+\bm{r}_2)$ and $\bm{r}=\bm{r}_1-\bm{r}_2$ are the center-of-mass coordinate and relative coordinate, respectively, and $P(r)$ reads
 \begin{equation}
 P(r) = \frac{1}{(4\pi a^2)^{3/2}} e^{-\frac{r^2}{4a^2}}.
 \end{equation}
The parameters of the separable force are determined by reproducing the pairing gap at the Fermi surface $\Delta(k_F)$ as a function of the Fermi momentum in nuclear matter calculated with the Gogny force D1S. This yields the values $G=728$ MeV fm$^3$ and $a=0.644$ fm~\cite{Tian2009}.

Using the solutions of the FTRHB equations, one can calculate the pairing energy,
\begin{equation}
    E_{\rm pair}  =  {\rm Tr}(\Delta \kappa )
    =   \sum_{ik} (\Delta_{ik} \kappa_{ik} ),
\end{equation}
and the averaged pairing gap,
   \begin{equation}
    \Delta  = \frac{ E_{\rm pair}  }{  {\rm Tr}\kappa}
    = \frac{  \sum_{ik} (\Delta_{ik} \kappa_{ik} )}{  \sum_k \kappa_{kk}}.
   \end{equation}
The entropy of the system is evaluated from
  \begin{equation}
   S(T)
   =- k_{\rm B} \sum_i \left[f_i{\rm ln}f_i + (1-f_i) {\rm ln}(1-f_i)  \right],
  \end{equation}
and the specific heat is defined by the relation
  \begin{equation}
  \label{sheat}
   C_{v}(T) = \partial E^* / \partial T,
  \end{equation}
where $E^* = E(T) - E(T=0)$ is the internal excitation energy.

The self-consistent FTRHB equations are solved in the spherical harmonic oscillator basis. In this study all the calculations are performed in a large basis of 20 major oscillator shells. The mean field is determined by the effective interaction PC-PK1~\cite{Zhao2010}. For the pairing interaction the matrix elements of the separable pairing force can be represented by a sum of a few separable terms in the basis of spherical harmonic oscillator functions~\cite{Tian2009}. In practical applications it has been found that this sum can be approximated by a finite value $N_0=8$.

\section{Results and discussions}
\label{sec3}

\begin{figure}
\centerline{
\includegraphics[scale=0.35,angle=0]{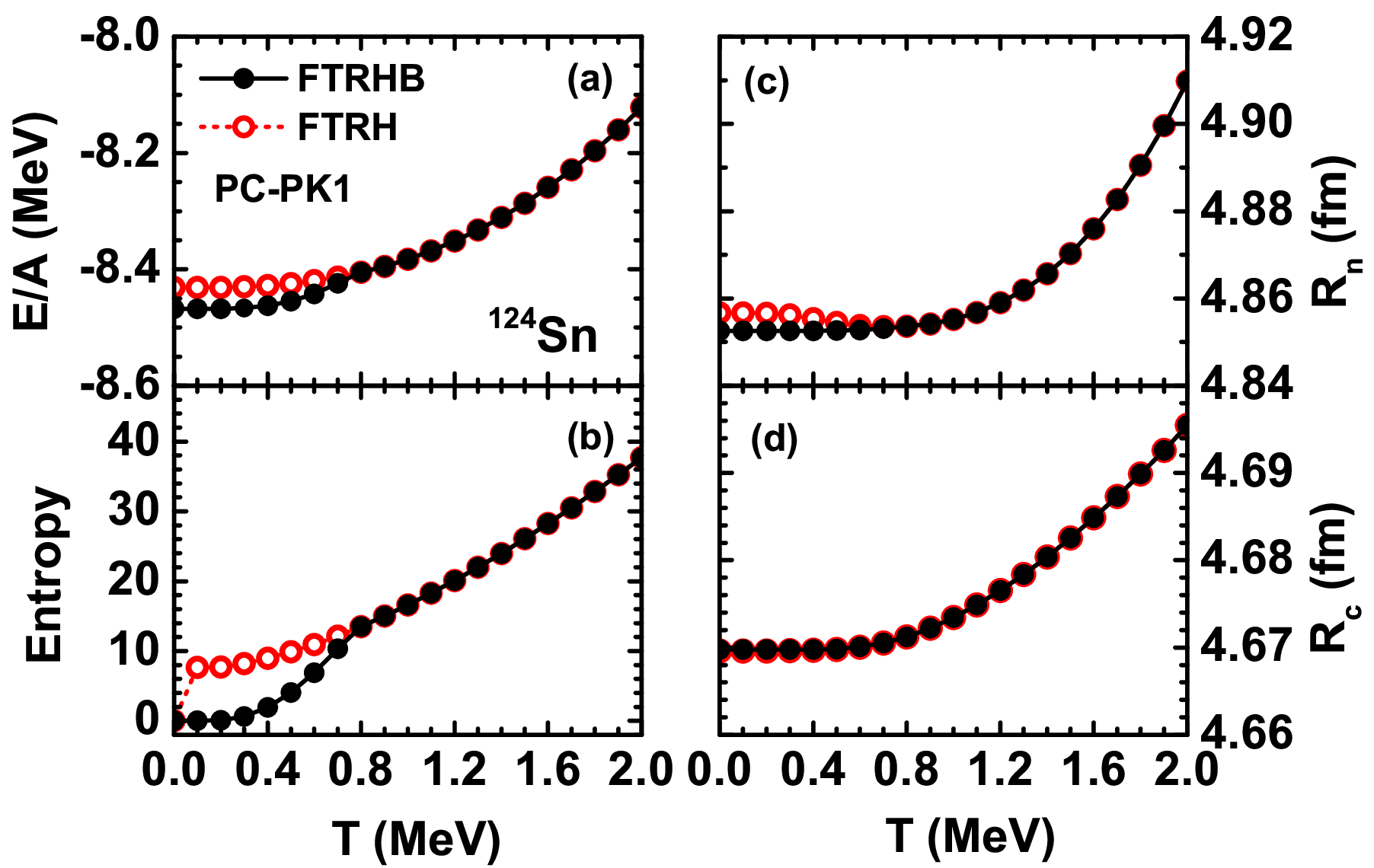}
} \caption{(Color online) Binding energy per nucleon (a), entropy (b), neutron radius (c), and charge radius (d) as a function of temperature $T$ for the nucleus $^{124}$Sn, calculated with the FTRH (red open circles) and FTRHB theories (black solid circles), using the effective interaction PC-PK1.}
\label{fig1}
\end{figure}

\begin{figure}
\centerline{
\includegraphics[scale=0.35,angle=0]{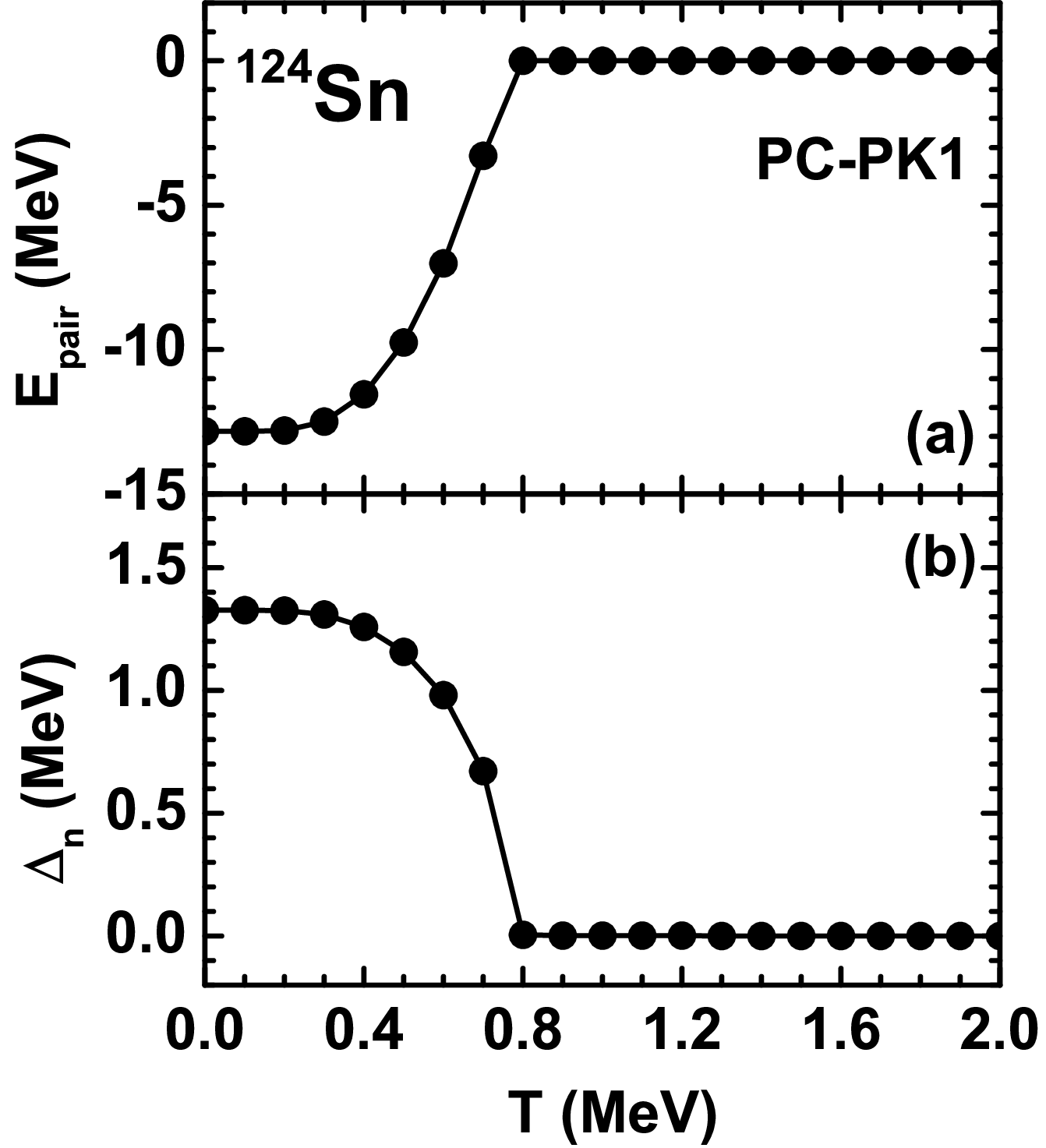}
} \caption{The neutron pairing energy (a) and the neutron pairing gap (b) for the nucleus $^{124}$Sn
as a function of temperature, calculated in the FTRHB theory with the effective interaction PC-PK1. }
\label{fig2}
\end{figure}

In Fig.~\ref{fig1} we display the binding energy per nucleon, entropy, neutron radius, and charge radius as a function of the temperature $T$ for the nucleus $^{124}$Sn, calculated using the effective interaction PC-PK1 in the FT relativistic Hartree (FTRH) theory without pairing correlations and the FTRHB theory with pairing correlations included. The relativistic Hartree theory is also known as the RMF theory. In the FTRHB calculation the Gogny force with the parametrization D1S is employed in the pairing channel.  Comparing the FTRH and FTRHB results, one notices that these two approaches yield the same results for the charge radius at all temperatures $T \leq 2$ MeV, because $^{124}$Sn is a semimagic nucleus with a magic proton number. For the other quantities with the contributions of neutrons such as the binding energy per nucleon, entropy and neutron radius, differences can be seen at low temperatures. However, the differences obtained with and without the inclusion of pairing correlations vanish at temperatures $T \geq 0.8$ MeV. Pairing correlations, therefore, no longer play a role beyond $T=0.8$ MeV, which implies that a transition from the superfluid phase to the normal one occurs at this critical temperature. This is further verified by the evolution of the neutron pairing energy and pairing gap with temperature in Fig.~\ref{fig2}. With the temperature increasing to $T=0.8$ MeV, the differences between the FTRH and the FTRHB results decrease gradually to zero as the correlated nucleon pairs are broken, and finally the normal phase without pairing correlations is reached.

In panel (a) of Fig.~\ref{fig1} the binding energy per nucleon increases quadratically with temperature after $T_c$. This is in accordance with the Fermi gas model, in which the temperature dependence of the excitation energy is $E^* = aT^2$, and $a$ is the level-density parameter~\cite{BohrMottelsonbook,Bethe1936,Sano1963}.
The binding energy per nucleon increases by 4\% from zero temperature to $T=2$ MeV, and correspondingly the excitation energy $E^*$ reaches $42.9$ MeV at $T=2$ MeV.
In panel (b) the entropy increases quadratically with temperature for $T < T_c$. The temperature dependence  becomes linear as soon as the transition to the normal phase occurs, as expected from the Fermi gas model in which $S=2aT$. In panels (c) and (d) the neutron radius and charge radius, respectively, show almost no variations until $T=0.8$ MeV and then begin to increase gradually. The neutron radius increases by 1\% from zero temperature to $T=2$ MeV, while for the charge radius the increase is only 0.6\% because protons are constrained by the Coulomb barrier.
Temperature increase leads to the excitations of individual nucleons to higher energy orbitals, including loosely bound levels and even the continuum. This causes a small increase of nuclear radii. At much higher temperatures, e.g., above 4 MeV, more nucleons enter the continuum, and the effect of the nucleon vapor needs to be taken into account~\cite{Bonche1984}. In the present study calculations are limited to the range $T \leq 2$ MeV and continuum contributions need not be considered.

To display the pairing transition more clearly, in Fig.~\ref{fig2} we plot the neutron pairing energy and the neutron pairing gap as functions of temperature for the nucleus $^{124}$Sn, calculated in the FTRHB theory with the effective interaction PC-PK1 and the Gogny pairing interaction D1S. The pairing energy does not vary much at low temperatures $T \leq 0.4$ MeV, but then it increases abruptly to zero at $T=0.8$ MeV. Correspondingly, the pairing gap displays a pronounced decrease starting from $T \approx 0.4$ MeV, and vanishes  at $T=0.8$ MeV. The critical temperature above which pairing correlations vanish, i.e., $T_c=0.8$ MeV, corresponds to $0.6$ times $\Delta_n(0)$, where $\Delta_n(0)=1.33$ MeV is the neutron pairing gap at zero temperature. The disappearance of the pairing energy and pairing gap at the critical temperature corresponds to a transition from the superfluid phase to the normal phase.

\begin{figure}
\centerline{
\includegraphics[scale=0.35,angle=0]{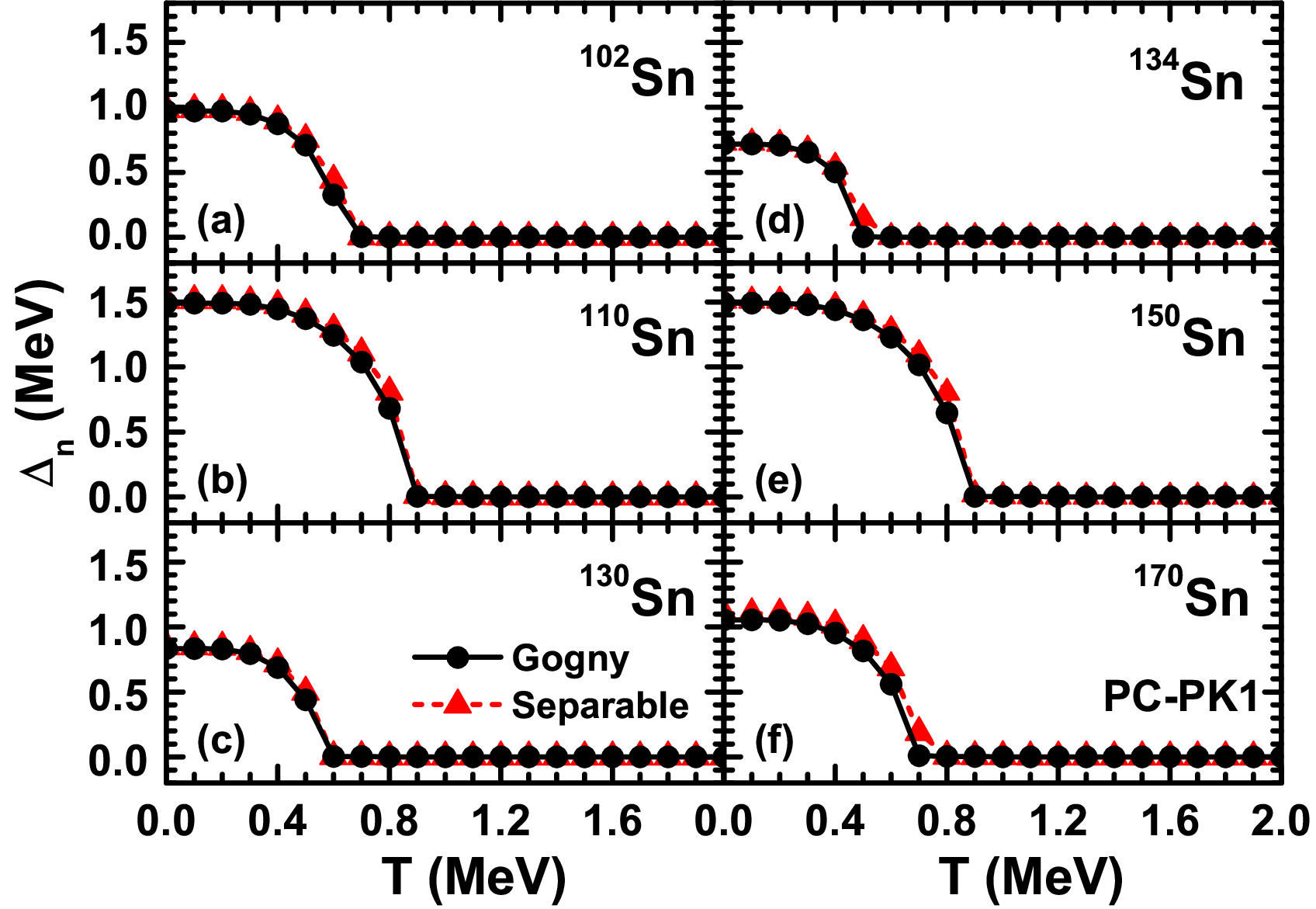}
} \caption{(Color online) The neutron pairing gap as a function of temperature for selected nuclei in the Sn isotopic chain, calculated in the  FTRHB theory with the Gogny pairing interaction D1S (black solid circles) and separable pairing interaction (red triangles). }
\label{fig3}
\end{figure}

To analyze the systematic evolution of pairing gaps with temperature, we select several Sn isotopes and plot the neutron pairing gaps as functions of temperature in Fig.~\ref{fig3}. The Gogny force and the separable pairing force are employed in the pairing channel of the FTRHB calculation.
In general, the separable pairing force reproduces the neutron pairing gaps calculated with the Gogny interaction, not only at zero temperature but also at finite temperatures for all nuclei analyzed here. The same temperature dependence of the pairing gaps is predicted by both pairing interactions. From the neutron-deficient $^{102}$Sn to the very neutron-rich $^{170}$Sn, the evolution of neutron pairing gaps with temperature follows a similar pattern,  rapidly decreasing as $T$ approaches the critical temperature. For the Sn isotopes shown in Fig.~\ref{fig3} the neutron number crosses two major shells. In the major shell $N=50-82$, the pairing gap at zero temperature varies from a relatively small value $\approx 1$ MeV for $^{102}$Sn with neutron number just beyond the shell closure at $N=50$, to a rather large value $\approx 1.5$ MeV for $^{110}$Sn with neutron number near midshell, and then again to $< 1$ MeV for $^{130}$Sn. In the major shell $N=82-126$ we have also selected three nuclei with neutron numbers near the shell closures and in the middle of the shell, and the same evolution of pairing gaps with neutron number is observed. Correspondingly, the critical temperature displays the same dependence on the neutron number as the pairing gaps. $T_c$ is relatively large for nuclei in the middle of the shell, and small for nuclei near neutron shell closure. Furthermore, it is found that the critical temperature follows very closely the relation $T_c = 0.6 \Delta_n(0)$, just as in the case of $^{124}$Sn, where $\Delta_n(0)$ is the neutron pairing gap at zero temperature. This is basically in accordance with the results of the BCS theory with a constant pairing strength $G$, where the critical temperature obeys the relation $T_c = 0.57 \Delta(0)$.

\begin{figure}
\centerline{
\includegraphics[scale=0.35,angle=0]{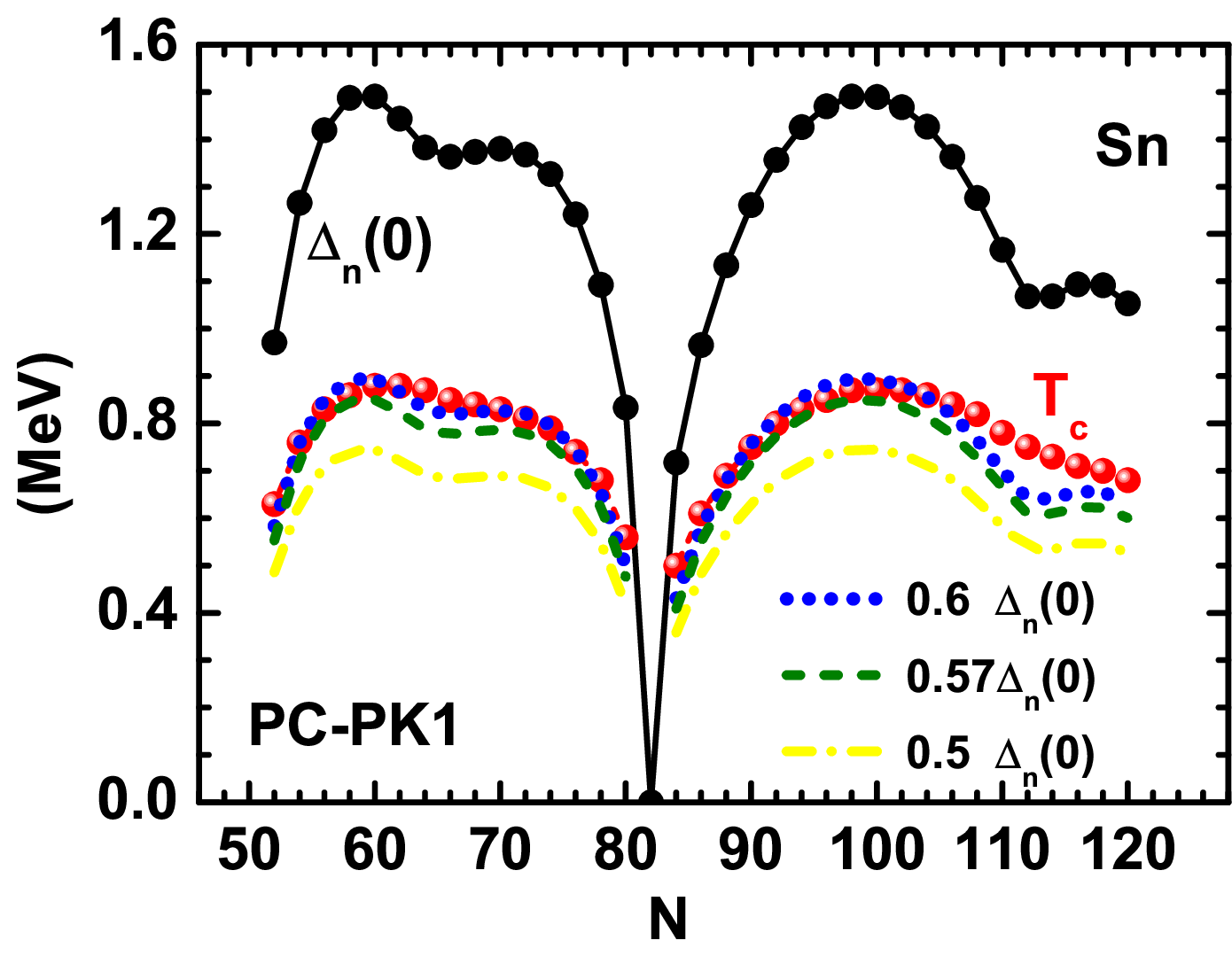}
} \caption{(Color online) The neutron pairing gaps at zero temperature (black solid circles) and the critical temperatures for pairing transition (red solid circles) in the even-even Sn isotopes, calculated in the FTRHB theory with the effective interaction PC-PK1 and the Gogny pairing interaction D1S. The scaled values of the neutron pairing gap at zero temperature: $0.6 \Delta_n(0)$, $0.57 \Delta_n(0)$, and $0.5 \Delta_n(0)$ are denoted by the blue (dotted), green (dashed), and yellow (dash-dotted) curves, respectively. }
\label{fig4}
\end{figure}

Figure \ref{fig4} displays the detailed isotopic dependence of the neutron pairing gaps at zero temperature and  the critical temperatures for pairing transition in even-even Sn nuclei, calculated using the FTRHB theory with the effective interaction PC-PK1 and the Gogny pairing interaction D1S. For comparison, the curves with the scaled values of the neutron pairing gap at zero temperature: $0.6 \Delta_n(0)$, $0.57 \Delta_n(0)$, and $0.5 \Delta_n(0)$ are also shown in the figure. Within a major shell, the pairing gap first increases as the neutron number approaches the middle
of the shell; then it decreases to zero at the neutron magic number. The critical temperature follows the same isotopic dependence and coincides very well with the curve $0.6 \Delta_n(0)$ for the whole isotopic chain, except for some very neutron-rich nuclei $^{160,162,164}$Sn. The largest discrepancy between the calculated critical temperature and the approximate empirical value $0.6 \Delta_n(0)$ is $0.11$ MeV for nucleus $^{162}$Sn. This is because for $N=112$ there is a subshell closure with the filling of the orbital  3$p_{1/2}$, and the gap between 3$p_{1/2}$ and 1$i_{13/2}$ is about 2.6 MeV. This subshell gap is not as large as a major shell gap so that the pairing gap for $^{162}$Sn does not vanish, but is considerably reduced compared to  neighboring nuclei. The occurrence of subshell structures indicates that a constant level density within a major shell is not a very good approximation, and this is reflected in the observed deviation from the simple relationship $0.6 \Delta_n(0)$ between the critical temperature and the zero-temperature pairing gap. In the vicinity of subshell closures the pairing gaps can reflect the underlying shell structures, whereas critical temperatures always display a smooth variation.

Through the comparison between the microscopic Bogoliubov model calculation and the theoretical value for critical temperature, we can see that $T_c=0.5 \Delta_n(0)$ calculated from the degenerate BCS model obviously underestimates the critical temperature, where working within the half-filled degenerate single-$j$ shell, which is similar as the seniority model, shows a relatively poor approximation. However, $T_c=0.57\Delta(0)$ obtained from the BCS theory gives a much better estimation of the critical temperature; however, it still slightly underestimates the critical temperature calculated from the Bogoliubov theory, and gives less accurate critical temperatures compared with $T_c=0.6\Delta(0)$. This underestimation may be attributable to the fact that in Bogoliubov theory, not only the state and its time-reversal state but also states from different single-particle levels could be paired, and hence more correlations are included, so higher temperature is required to break all the paired states.

\begin{figure}
\centerline{
\includegraphics[scale=0.22,angle=0]{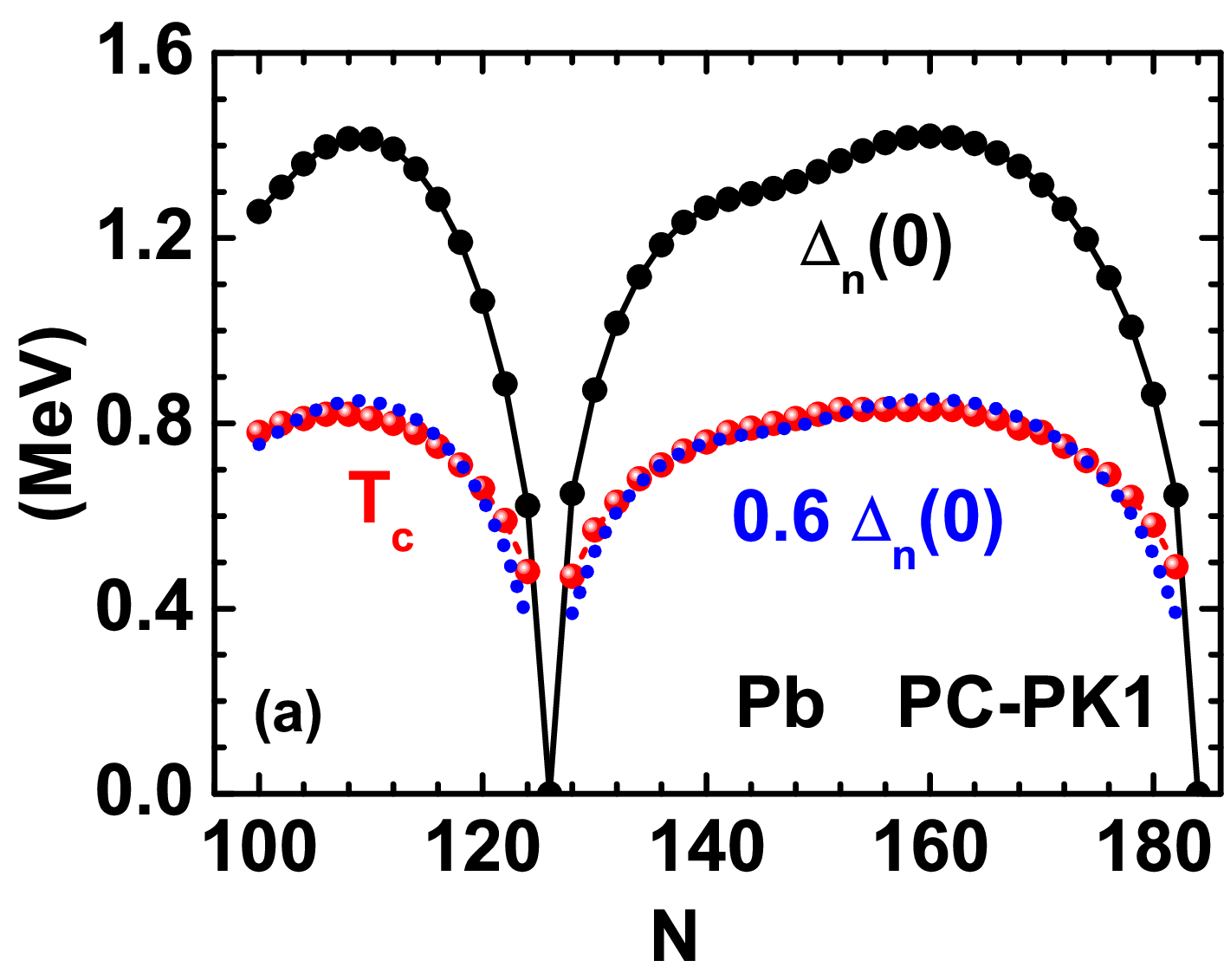}
\includegraphics[scale=0.22,angle=0]{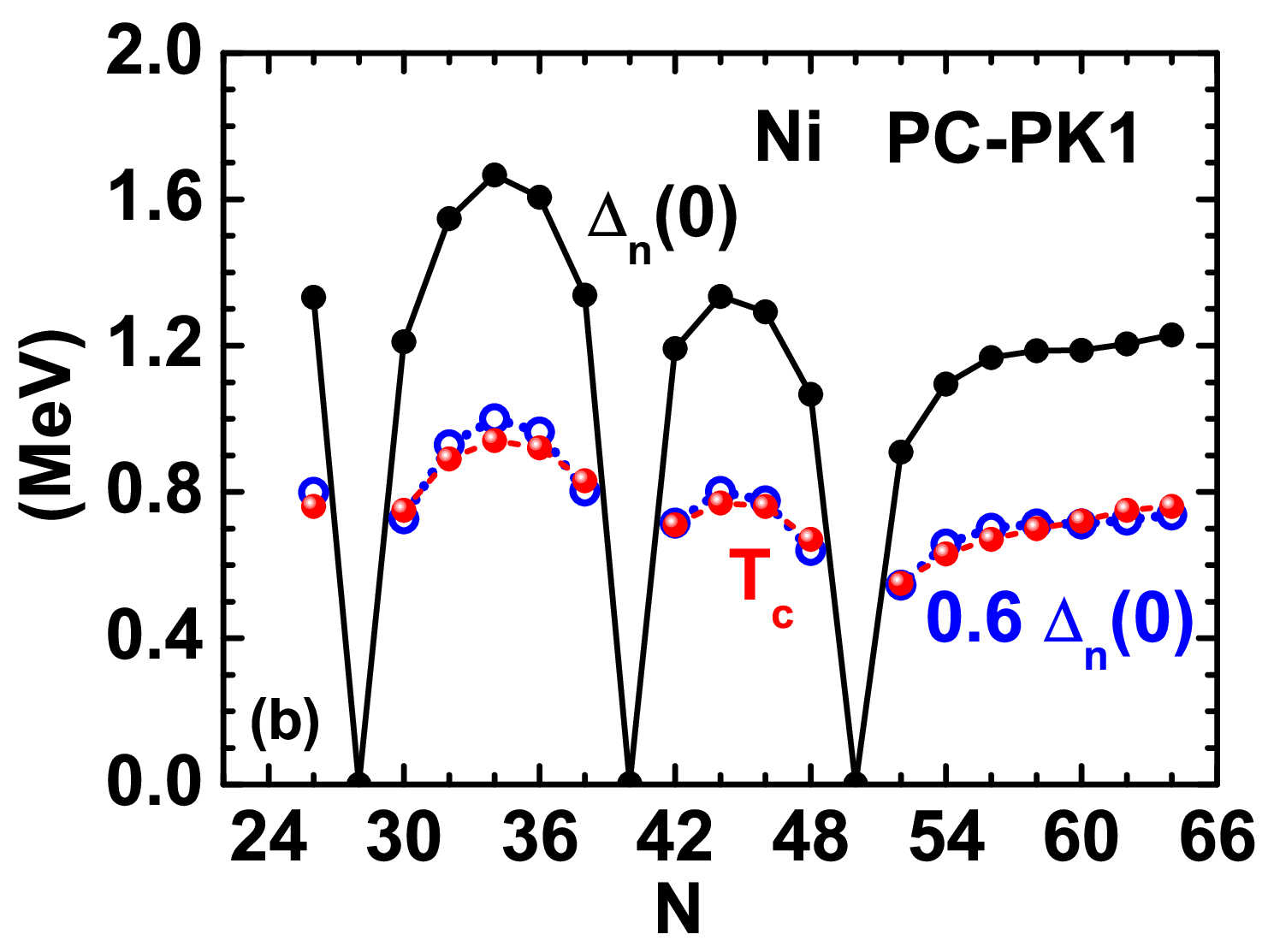}
\includegraphics[scale=0.22,angle=0]{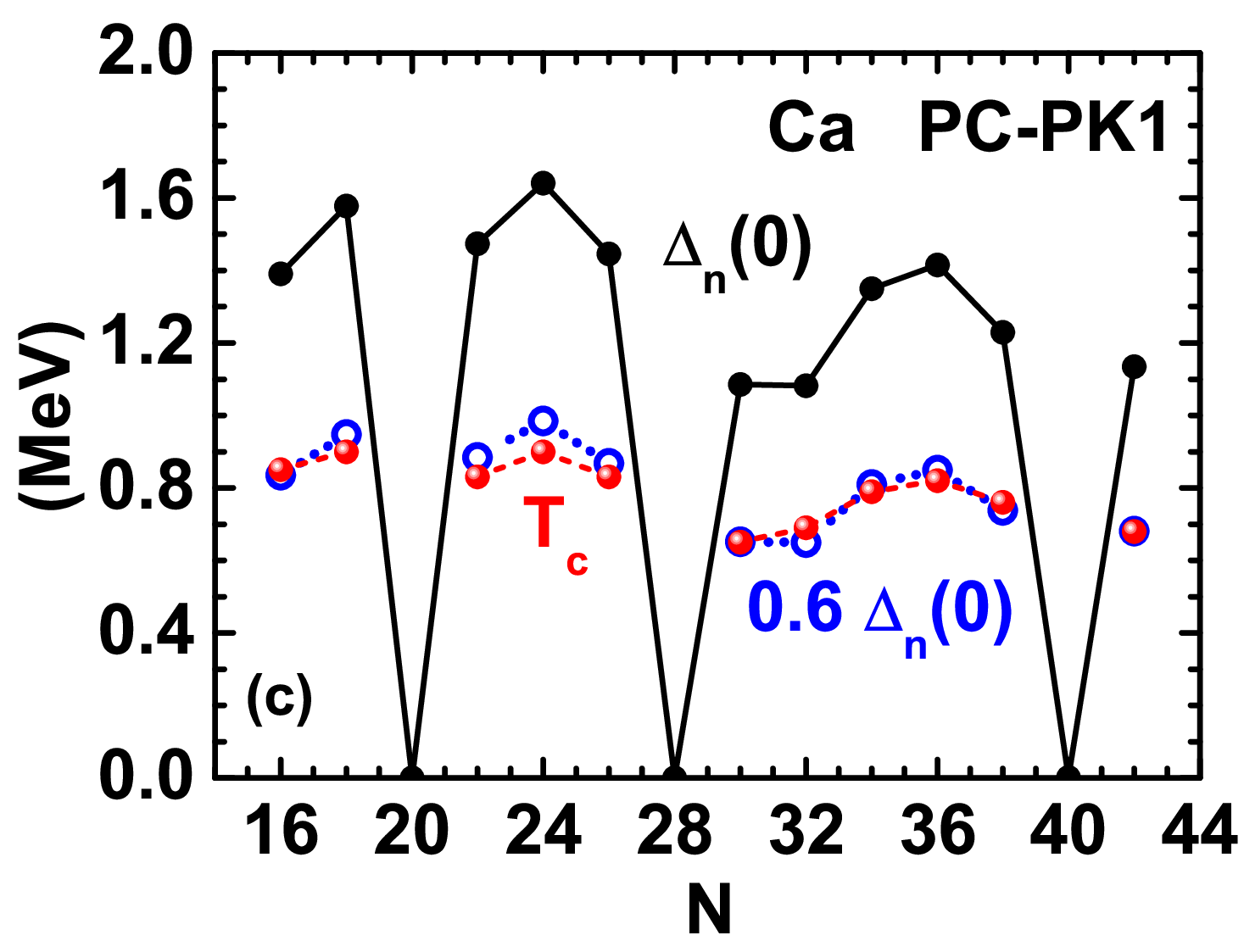}
} \caption{(Color online) Same as in the caption to Fig.~\ref{fig4} but for Pb (a), Ni (b), and Ca isotopes (c). }
\label{fig5}
\end{figure}

For Sn isotopes with the valence neutrons spanning the two major shells $N=50-82$ and $N=82-126$, the critical temperature closely follows the curve $T_c = 0.6 \Delta(0)$. To verify that this dependence is universal for other isotopic chains, and with valence nucleons in other major shells, in Fig.~\ref{fig5} we plot the neutron pairing gaps at zero temperature and the critical temperatures for the even-even Pb, Ni and Ca isotopes. For the Pb isotopic chain the valence neutrons occupy part of the major shell $N=82-126$, and the whole shell $N=126-184$. For the Ni isotopes the valence neutrons span the major shell $N=28-50$, and occupy part of the shell $N=50-82$. For the Ca isotopes the major shells $N=8-20$ and $N=20-50$ are occupied by valence neutrons. Despite the large interval of occupied valence shells, in all three cases the isotopic dependence of the critical temperature calculated using the FTRHB theory is accurately reproduced by the relation  $T_c = 0.6 \Delta(0)$.

\begin{figure}
\centerline{
\includegraphics[scale=0.35,angle=0]{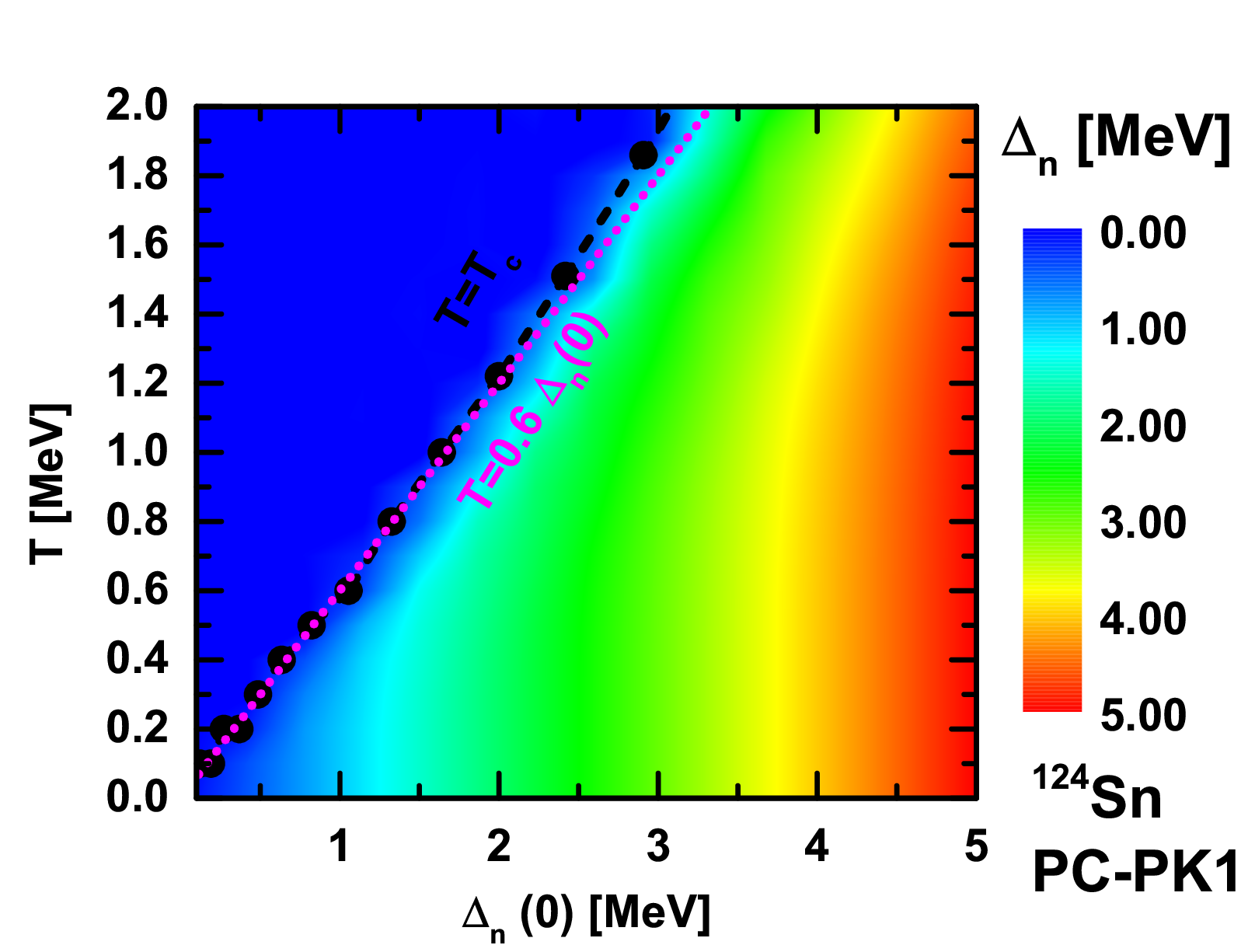}
} \caption{(Color online) Contour plot for the neutron pairing gap in $^{124}$Sn as a function of the temperature and neutron pairing gap at zero temperature. The FTRHB calculation is carried out with the effective interaction PC-PK1 and the Gogny pairing interaction. The strength parameter of the latter is varied to obtain different values of the pairing gap at zero temperature. For each value of the zero-temperature pairing gap the critical temperature $T_c$ is denoted by a black solid circle, and the dotted line corresponds to the relation $T_c = 0.6 \Delta(0)$.}
\label{fig6}
\end{figure}

To test the rule $T_c=0.6\Delta(0)$ in a different way, in Fig.~\ref{fig6} we display the contour plot for the neutron pairing gap in $^{124}$Sn as a function of the temperature and neutron pairing gap at zero temperature. The latter is varied by changing the strength of the Gogny pairing interaction in the interval $0.1$ to $1.8$ of the original value. The corresponding critical temperatures and the linear relation $0.6 \Delta_n(0)$ are also shown in the figure.
It is interesting to note that the critical temperature closely follows the relation $T_c=0.6\Delta(0)$ over a wide range of zero-temperature pairing gaps. $T_c$ only starts to deviate from the line $0.6 \Delta_n(0)$, and higher values of the critical temperature are obtained, when the neutron pairing gap at zero temperature exceeds 2.5 MeV. Therefore, we conclude that the relation $T_c=0.6\Delta(0)$ holds over a wide interval of the pairing strength parameters and the corresponding neutron pairing gaps at zero temperature.

\begin{figure}
\centerline{
\includegraphics[scale=0.35,angle=0]{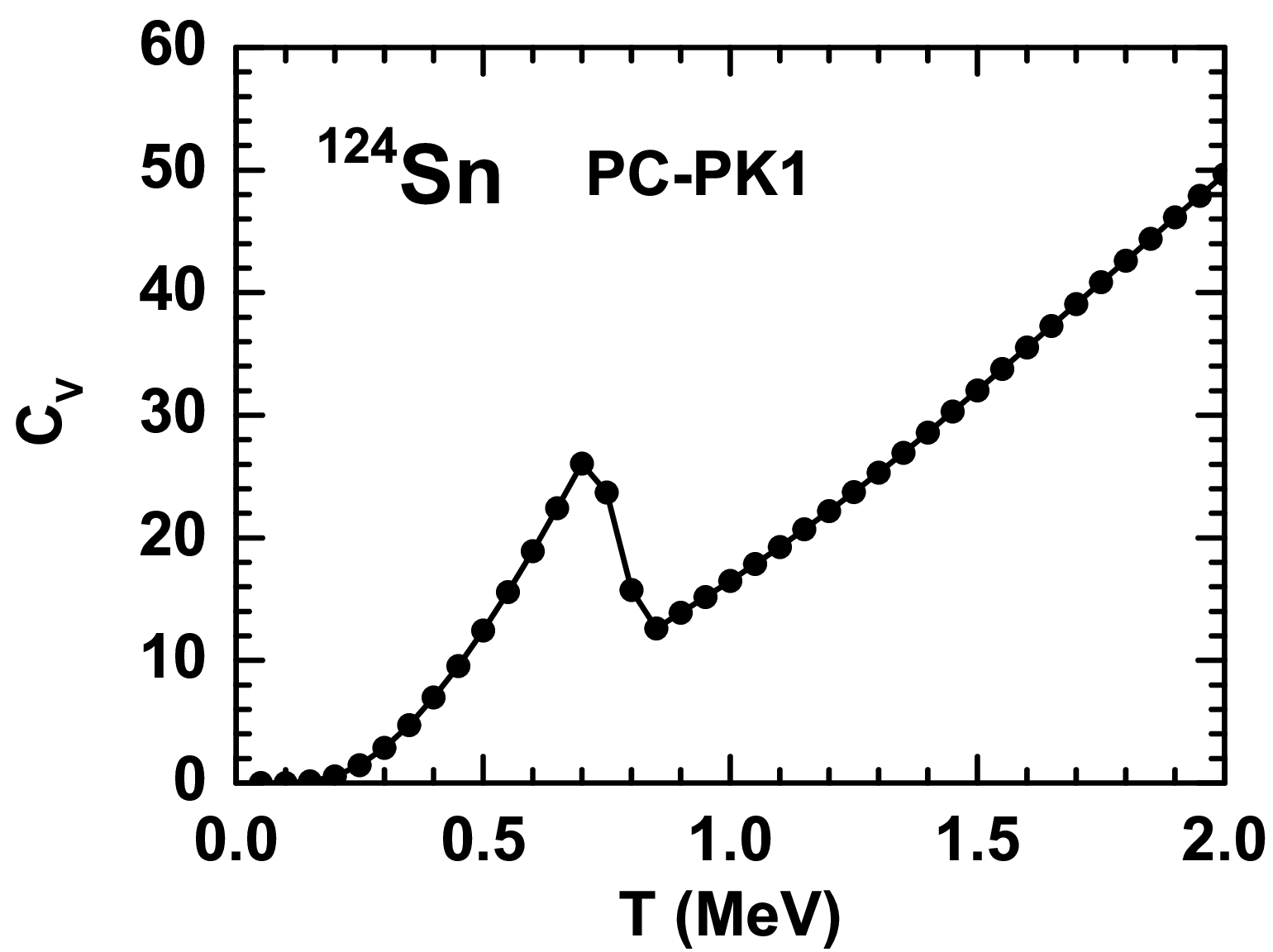}
} \caption{Specific heat for the nucleus $^{124}$Sn as a function of temperature $T$, calculated using the FTRHB with the effective interaction PC-PK1 and the Gogny pairing force.}
\label{fig7}
\end{figure}

Finally, the specific heat for the nucleus $^{124}$Sn as a function of temperature, calculated with effective interaction PC-PK1 and Gogny pairing force D1S, is shown in Fig.~\ref{fig7}. The marked discontinuity at the critical temperature indicates the transition from the superfluid to the normal phase. At low temperature the increase of the specific heat with temperature is nonlinear, whereas a linear increase is calculated
for temperatures beyond $T_c$. However, in experiment the specific heat usually exhibits a more smooth S-shaped behavior close to the critical temperature, as compared to the sharp discontinuity obtained in the present  calculation. This is attributable to the finite size of the nucleus and, therefore, thermal fluctuations need to be taken into account for a more realistic description.

\section{Conclusion}
\label{sec4}

The FT relativistic Hartree-Bogoliubov theory for spherical nuclei, based on the point-coupling functional PC-PK1 and with the Gogny or separable pairing forces, has been formulated and applied to the study of pairing transition in Ca, Ni, Sn and Pb isotopes. The FTRHB theory coincides with the FTRH theory in the description of physical quantities such as the binding energy per nucleon, the radius and the entropy once the transition from the superfluid phase to the normal one occurs. It is found that the separable pairing force reproduces the pairing gaps calculated using the Gogny force not only at zero temperature, but also at finite temperatures. The same evolution of pairing gaps with temperature is found for both interactions. In a detailed calculation of even-even isotopes of Ca, Ni, Sn, and Pb, it is found that the critical temperature for the pairing transition closely follows the linear relation $T_c =0.6 \Delta(0)$, even with valence neutrons spanning several major shells. This rule has been further verified by varying  the pairing gap at zero temperature using different values of the pairing strength parameter.

The formulation of the FTRHB theory provides a basis for the further development of the FT relativistic quasiparticle random phase approximation (FTRQRPA), which is a powerful tool for the self-consistent description of electron capture and $\beta$ decay in stellar environment. Although stellar electron capture has been investigated using the FT random phase approximation based on energy density functionals, these calculations have not included pairing correlations so far. Stellar temperatures at which electron capture and, especially, $\beta$ decays take place can be below the critical temperature for pairing phase transition. Therefore, for such processes pairing correlations might still play an important role, and the development of the FTRQRPA will be necessary for more accurate descriptions of stellar weak-interaction processes. In the FTRQRPA model one needs to work in the quasiparticle basis instead of the usual canonical basis, and the use of point-coupling functionals and separable pairing forces, introduced in the present work, simplifies the implementation of this model.
{\center{\bf ACKNOWLEDGMENTS}}

Y. F. Niu would like to acknowledge discussions with S. Chen. This work was supported in part by the Major State 973 Program 2013CB834400, the NSFC under Grants No. 11175002, No. 11205004, No. 11305161, No. 11072228, No. 11372294, No. 11172277 and the MZOS-project No. 1191005-1010.


%

\clearpage

\begin{thebibliography}{68}
\expandafter\ifx\csname natexlab\endcsname\relax\def\natexlab#1{#1}\fi
\expandafter\ifx\csname bibnamefont\endcsname\relax
  \def\bibnamefont#1{#1}\fi
\expandafter\ifx\csname bibfnamefont\endcsname\relax
  \def\bibfnamefont#1{#1}\fi
\expandafter\ifx\csname citenamefont\endcsname\relax
  \def\citenamefont#1{#1}\fi
\expandafter\ifx\csname url\endcsname\relax
  \def\url#1{\texttt{#1}}\fi
\expandafter\ifx\csname urlprefix\endcsname\relax\def\urlprefix{URL }\fi
\providecommand{\bibinfo}[2]{#2}
\providecommand{\eprint}[2][]{\url{#2}}

\bibitem[{\citenamefont{Bardeen et~al.}(1957)\citenamefont{Bardeen, Cooper, and
  Schrieffer}}]{Bardeen1957}
\bibinfo{author}{\bibfnamefont{J.}~\bibnamefont{Bardeen}},
  \bibinfo{author}{\bibfnamefont{L.~N.} \bibnamefont{Cooper}},
  \bibnamefont{and} \bibinfo{author}{\bibfnamefont{J.~R.}
  \bibnamefont{Schrieffer}}, \bibinfo{journal}{Phys. Rev.}
  \textbf{\bibinfo{volume}{108}}, \bibinfo{pages}{1175} (\bibinfo{year}{1957}).

\bibitem[{\citenamefont{Langanke et~al.}(2005)\citenamefont{Langanke, Dean, and
  Nazarewicz}}]{Langanke2005}
\bibinfo{author}{\bibfnamefont{K.}~\bibnamefont{Langanke}},
  \bibinfo{author}{\bibfnamefont{D.~J.} \bibnamefont{Dean}}, \bibnamefont{and}
  \bibinfo{author}{\bibfnamefont{W.}~\bibnamefont{Nazarewicz}},
  \bibinfo{journal}{Nucl. Phys. A} \textbf{\bibinfo{volume}{757}},
  \bibinfo{pages}{360} (\bibinfo{year}{2005}).

\bibitem[{\citenamefont{M\"ulken et~al.}(2001)\citenamefont{M\"ulken,
  Stamerjohanns, and Borrmann}}]{Mulken2001}
\bibinfo{author}{\bibfnamefont{O.}~\bibnamefont{M\"ulken}},
  \bibinfo{author}{\bibfnamefont{H.}~\bibnamefont{Stamerjohanns}},
  \bibnamefont{and} \bibinfo{author}{\bibfnamefont{P.}~\bibnamefont{Borrmann}},
  \bibinfo{journal}{Phys. Rev. E} \textbf{\bibinfo{volume}{64}},
  \bibinfo{pages}{047105} (\bibinfo{year}{2001}).

\bibitem[{\citenamefont{Egido et~al.}(2000)\citenamefont{Egido, Robledo, and
  Martin}}]{Egido2000}
\bibinfo{author}{\bibfnamefont{J.~L.} \bibnamefont{Egido}},
  \bibinfo{author}{\bibfnamefont{L.~M.} \bibnamefont{Robledo}},
  \bibnamefont{and} \bibinfo{author}{\bibfnamefont{V.}~\bibnamefont{Martin}},
  \bibinfo{journal}{Phys. Rev. Lett.} \textbf{\bibinfo{volume}{85}},
  \bibinfo{pages}{26} (\bibinfo{year}{2000}).

\bibitem[{\citenamefont{Melby et~al.}(1999)\citenamefont{Melby, Bergholt,
  Guttormsen, Hjorth-Jensen, Ingebretsen, Messelt, Rekstad, Schiller, Siem, and
  \O{}deg\aa{}rd}}]{Melby1999}
\bibinfo{author}{\bibfnamefont{E.}~\bibnamefont{Melby}},
  \bibinfo{author}{\bibfnamefont{L.}~\bibnamefont{Bergholt}},
  \bibinfo{author}{\bibfnamefont{M.}~\bibnamefont{Guttormsen}},
  \bibinfo{author}{\bibfnamefont{M.}~\bibnamefont{Hjorth-Jensen}},
  \bibinfo{author}{\bibfnamefont{F.}~\bibnamefont{Ingebretsen}},
  \bibinfo{author}{\bibfnamefont{S.}~\bibnamefont{Messelt}},
  \bibinfo{author}{\bibfnamefont{J.}~\bibnamefont{Rekstad}},
  \bibinfo{author}{\bibfnamefont{A.}~\bibnamefont{Schiller}},
  \bibinfo{author}{\bibfnamefont{S.}~\bibnamefont{Siem}}, \bibnamefont{and}
  \bibinfo{author}{\bibfnamefont{S.~W.} \bibnamefont{\O{}deg\aa{}rd}},
  \bibinfo{journal}{Phys. Rev. Lett.} \textbf{\bibinfo{volume}{83}},
  \bibinfo{pages}{3150} (\bibinfo{year}{1999}).

\bibitem[{\citenamefont{Schiller et~al.}(2001)\citenamefont{Schiller, Bjerve,
  Guttormsen, Hjorth-Jensen, Ingebretsen, Melby, Messelt, Rekstad, Siem, and
  \O{}deg\aa{}rd}}]{Schiller2001}
\bibinfo{author}{\bibfnamefont{A.}~\bibnamefont{Schiller}},
  \bibinfo{author}{\bibfnamefont{A.}~\bibnamefont{Bjerve}},
  \bibinfo{author}{\bibfnamefont{M.}~\bibnamefont{Guttormsen}},
  \bibinfo{author}{\bibfnamefont{M.}~\bibnamefont{Hjorth-Jensen}},
  \bibinfo{author}{\bibfnamefont{F.}~\bibnamefont{Ingebretsen}},
  \bibinfo{author}{\bibfnamefont{E.}~\bibnamefont{Melby}},
  \bibinfo{author}{\bibfnamefont{S.}~\bibnamefont{Messelt}},
  \bibinfo{author}{\bibfnamefont{J.}~\bibnamefont{Rekstad}},
  \bibinfo{author}{\bibfnamefont{S.}~\bibnamefont{Siem}}, \bibnamefont{and}
  \bibinfo{author}{\bibfnamefont{S.~W.} \bibnamefont{\O{}deg\aa{}rd}},
  \bibinfo{journal}{Phys. Rev. C} \textbf{\bibinfo{volume}{63}},
  \bibinfo{pages}{021306} (\bibinfo{year}{2001}).

\bibitem[{\citenamefont{Melby et~al.}(2001)\citenamefont{Melby, Guttormsen,
  Rekstad, Schiller, Siem, and Voinov}}]{Melby2001}
\bibinfo{author}{\bibfnamefont{E.}~\bibnamefont{Melby}},
  \bibinfo{author}{\bibfnamefont{M.}~\bibnamefont{Guttormsen}},
  \bibinfo{author}{\bibfnamefont{J.}~\bibnamefont{Rekstad}},
  \bibinfo{author}{\bibfnamefont{A.}~\bibnamefont{Schiller}},
  \bibinfo{author}{\bibfnamefont{S.}~\bibnamefont{Siem}}, \bibnamefont{and}
  \bibinfo{author}{\bibfnamefont{A.}~\bibnamefont{Voinov}},
  \bibinfo{journal}{Phys. Rev. C} \textbf{\bibinfo{volume}{63}},
  \bibinfo{pages}{044309} (\bibinfo{year}{2001}).

\bibitem[{\citenamefont{Guttormsen et~al.}(2003)\citenamefont{Guttormsen,
  Chankova, Hjorth-Jensen, Rekstad, Siem, Schiller, and Dean}}]{Guttormsen2003}
\bibinfo{author}{\bibfnamefont{M.}~\bibnamefont{Guttormsen}},
  \bibinfo{author}{\bibfnamefont{R.}~\bibnamefont{Chankova}},
  \bibinfo{author}{\bibfnamefont{M.}~\bibnamefont{Hjorth-Jensen}},
  \bibinfo{author}{\bibfnamefont{J.}~\bibnamefont{Rekstad}},
  \bibinfo{author}{\bibfnamefont{S.}~\bibnamefont{Siem}},
  \bibinfo{author}{\bibfnamefont{A.}~\bibnamefont{Schiller}}, \bibnamefont{and}
  \bibinfo{author}{\bibfnamefont{D.~J.} \bibnamefont{Dean}},
  \bibinfo{journal}{Phys. Rev. C} \textbf{\bibinfo{volume}{68}},
  \bibinfo{pages}{034311} (\bibinfo{year}{2003}).

\bibitem[{\citenamefont{Sano and Yamasaki}(1963)}]{Sano1963}
\bibinfo{author}{\bibfnamefont{M.}~\bibnamefont{Sano}} \bibnamefont{and}
  \bibinfo{author}{\bibfnamefont{S.}~\bibnamefont{Yamasaki}},
  \bibinfo{journal}{Prog. Theo. Phys.} \textbf{\bibinfo{volume}{29}},
  \bibinfo{pages}{397} (\bibinfo{year}{1963}).

\bibitem[{\citenamefont{Goodman}(1981)}]{Goodman1981}
\bibinfo{author}{\bibfnamefont{A.~L.} \bibnamefont{Goodman}},
  \bibinfo{journal}{Nucl. Phys. A} \textbf{\bibinfo{volume}{352}},
  \bibinfo{pages}{30} (\bibinfo{year}{1981}).

\bibitem[{\citenamefont{Levit and Alhassid}(1984)}]{Levit1984}
\bibinfo{author}{\bibfnamefont{S.}~\bibnamefont{Levit}} \bibnamefont{and}
  \bibinfo{author}{\bibfnamefont{Y.}~\bibnamefont{Alhassid}},
  \bibinfo{journal}{Nucl. Phys. A} \textbf{\bibinfo{volume}{413}},
  \bibinfo{pages}{439 } (\bibinfo{year}{1984}).

\bibitem[{\citenamefont{Goodman}(1984)}]{Goodman1984}
\bibinfo{author}{\bibfnamefont{A.~L.} \bibnamefont{Goodman}},
  \bibinfo{journal}{Phys. Rev. C} \textbf{\bibinfo{volume}{29}},
  \bibinfo{pages}{1887} (\bibinfo{year}{1984}).

\bibitem[{\citenamefont{Dang et~al.}(1993)\citenamefont{Dang, Ring, and
  Rossignoli}}]{Dang1993}
\bibinfo{author}{\bibfnamefont{N.~D.} \bibnamefont{Dang}},
  \bibinfo{author}{\bibfnamefont{P.}~\bibnamefont{Ring}}, \bibnamefont{and}
  \bibinfo{author}{\bibfnamefont{R.}~\bibnamefont{Rossignoli}},
  \bibinfo{journal}{Phys. Rev. C} \textbf{\bibinfo{volume}{47}},
  \bibinfo{pages}{606} (\bibinfo{year}{1993}).

\bibitem[{\citenamefont{Alhassid and Zingman}(1984)}]{Alhassid1984}
\bibinfo{author}{\bibfnamefont{Y.}~\bibnamefont{Alhassid}} \bibnamefont{and}
  \bibinfo{author}{\bibfnamefont{J.}~\bibnamefont{Zingman}},
  \bibinfo{journal}{Phys. Rev. C} \textbf{\bibinfo{volume}{30}},
  \bibinfo{pages}{684} (\bibinfo{year}{1984}).

\bibitem[{\citenamefont{Rossignoli and Ring}(1994)}]{Rossignoli1994}
\bibinfo{author}{\bibfnamefont{R.}~\bibnamefont{Rossignoli}} \bibnamefont{and}
  \bibinfo{author}{\bibfnamefont{P.}~\bibnamefont{Ring}},
  \bibinfo{journal}{Ann. Phys.} \textbf{\bibinfo{volume}{235}},
  \bibinfo{pages}{350 } (\bibinfo{year}{1994}).

\bibitem[{\citenamefont{Rossignoli et~al.}(1996)\citenamefont{Rossignoli,
  Canosa, and Egido}}]{Rossignoli1996}
\bibinfo{author}{\bibfnamefont{R.}~\bibnamefont{Rossignoli}},
  \bibinfo{author}{\bibfnamefont{N.}~\bibnamefont{Canosa}}, \bibnamefont{and}
  \bibinfo{author}{\bibfnamefont{J.}~\bibnamefont{Egido}},
  \bibinfo{journal}{Nucl. Phys. A} \textbf{\bibinfo{volume}{605}},
  \bibinfo{pages}{1 } (\bibinfo{year}{1996}).

\bibitem[{\citenamefont{Lang et~al.}(1993)\citenamefont{Lang, Johnson, Koonin,
  and Ormand}}]{Lang1993}
\bibinfo{author}{\bibfnamefont{G.~H.} \bibnamefont{Lang}},
  \bibinfo{author}{\bibfnamefont{C.~W.} \bibnamefont{Johnson}},
  \bibinfo{author}{\bibfnamefont{S.~E.} \bibnamefont{Koonin}},
  \bibnamefont{and} \bibinfo{author}{\bibfnamefont{W.~E.}
  \bibnamefont{Ormand}}, \bibinfo{journal}{Phys. Rev. C}
  \textbf{\bibinfo{volume}{48}}, \bibinfo{pages}{1518} (\bibinfo{year}{1993}).

\bibitem[{\citenamefont{Dean et~al.}(1995)\citenamefont{Dean, Koonin, Langanke,
  Radha, and Alhassid}}]{Dean1995}
\bibinfo{author}{\bibfnamefont{D.~J.} \bibnamefont{Dean}},
  \bibinfo{author}{\bibfnamefont{S.~E.} \bibnamefont{Koonin}},
  \bibinfo{author}{\bibfnamefont{K.}~\bibnamefont{Langanke}},
  \bibinfo{author}{\bibfnamefont{P.~B.} \bibnamefont{Radha}}, \bibnamefont{and}
  \bibinfo{author}{\bibfnamefont{Y.}~\bibnamefont{Alhassid}},
  \bibinfo{journal}{Phys. Rev. Lett.} \textbf{\bibinfo{volume}{74}},
  \bibinfo{pages}{2909} (\bibinfo{year}{1995}).

\bibitem[{\citenamefont{Langanke et~al.}(1996)\citenamefont{Langanke, Dean,
  Radha, and Koonin}}]{Langanke1996}
\bibinfo{author}{\bibfnamefont{K.}~\bibnamefont{Langanke}},
  \bibinfo{author}{\bibfnamefont{D.}~\bibnamefont{Dean}},
  \bibinfo{author}{\bibfnamefont{P.}~\bibnamefont{Radha}}, \bibnamefont{and}
  \bibinfo{author}{\bibfnamefont{S.}~\bibnamefont{Koonin}},
  \bibinfo{journal}{Nucl. Phys. A} \textbf{\bibinfo{volume}{602}},
  \bibinfo{pages}{244 } (\bibinfo{year}{1996}).

\bibitem[{\citenamefont{Liu and Alhassid}(2001)}]{Liu2001}
\bibinfo{author}{\bibfnamefont{S.}~\bibnamefont{Liu}} \bibnamefont{and}
  \bibinfo{author}{\bibfnamefont{Y.}~\bibnamefont{Alhassid}},
  \bibinfo{journal}{Phys. Rev. Lett.} \textbf{\bibinfo{volume}{87}},
  \bibinfo{pages}{022501} (\bibinfo{year}{2001}).

\bibitem[{\citenamefont{Kaneko and Hasegawa}(2004)}]{Kaneko2004}
\bibinfo{author}{\bibfnamefont{K.}~\bibnamefont{Kaneko}} \bibnamefont{and}
  \bibinfo{author}{\bibfnamefont{M.}~\bibnamefont{Hasegawa}},
  \bibinfo{journal}{Nucl. Phys. A} \textbf{\bibinfo{volume}{740}},
  \bibinfo{pages}{95} (\bibinfo{year}{2004}).

\bibitem[{\citenamefont{Bozzolo and Civitarese}(1985)}]{Bozzolo1985}
\bibinfo{author}{\bibfnamefont{G.}~\bibnamefont{Bozzolo}} \bibnamefont{and}
  \bibinfo{author}{\bibfnamefont{O.}~\bibnamefont{Civitarese}},
  \bibinfo{journal}{Phys. Rev. C} \textbf{\bibinfo{volume}{32}},
  \bibinfo{pages}{2111} (\bibinfo{year}{1985}).

\bibitem[{\citenamefont{Egido and Ring}(1993)}]{Egido1993}
\bibinfo{author}{\bibfnamefont{J.~L.} \bibnamefont{Egido}} \bibnamefont{and}
  \bibinfo{author}{\bibfnamefont{P.}~\bibnamefont{Ring}}, \bibinfo{journal}{J.
  Phys. G Nucl. Part. Phys.} \textbf{\bibinfo{volume}{19}}, \bibinfo{pages}{1}
  (\bibinfo{year}{1993}).

\bibitem[{\citenamefont{Goodman}(1986)}]{Goodman1986}
\bibinfo{author}{\bibfnamefont{A.~L.} \bibnamefont{Goodman}},
  \bibinfo{journal}{Phys. Rev. C} \textbf{\bibinfo{volume}{34}},
  \bibinfo{pages}{1942} (\bibinfo{year}{1986}).

\bibitem[{\citenamefont{Rei{\ss} et~al.}(1999)\citenamefont{Rei{\ss}, Bender,
  and Reinhard}}]{Reis1999}
\bibinfo{author}{\bibfnamefont{C.}~\bibnamefont{Rei{\ss}}},
  \bibinfo{author}{\bibfnamefont{M.}~\bibnamefont{Bender}}, \bibnamefont{and}
  \bibinfo{author}{\bibfnamefont{P.-G.} \bibnamefont{Reinhard}},
  \bibinfo{journal}{Eur. Phys. J. A.} \textbf{\bibinfo{volume}{6}},
  \bibinfo{pages}{157} (\bibinfo{year}{1999}).

\bibitem[{\citenamefont{Martin et~al.}(2003)\citenamefont{Martin, Egido, and
  Robledo}}]{Martin2003}
\bibinfo{author}{\bibfnamefont{V.}~\bibnamefont{Martin}},
  \bibinfo{author}{\bibfnamefont{J.~L.} \bibnamefont{Egido}}, \bibnamefont{and}
  \bibinfo{author}{\bibfnamefont{L.~M.} \bibnamefont{Robledo}},
  \bibinfo{journal}{Phys. Rev. C} \textbf{\bibinfo{volume}{68}},
  \bibinfo{pages}{034327} (\bibinfo{year}{2003}).

\bibitem[{\citenamefont{Khan et~al.}(2004)\citenamefont{Khan, Giai, and
  Grasso}}]{Khan2004}
\bibinfo{author}{\bibfnamefont{E.}~\bibnamefont{Khan}},
  \bibinfo{author}{\bibfnamefont{N.~Van} \bibnamefont{Giai}}, \bibnamefont{and}
  \bibinfo{author}{\bibfnamefont{M.}~\bibnamefont{Grasso}},
  \bibinfo{journal}{Nucl. Phys. A} \textbf{\bibinfo{volume}{731}},
  \bibinfo{pages}{311 } (\bibinfo{year}{2004}).

\bibitem[{\citenamefont{Khan et~al.}(2007)\citenamefont{Khan, Giai, and
  Sandulescu}}]{Khan2007}
\bibinfo{author}{\bibfnamefont{E.}~\bibnamefont{Khan}},
  \bibinfo{author}{\bibfnamefont{N.~Van} \bibnamefont{Giai}}, \bibnamefont{and}
  \bibinfo{author}{\bibfnamefont{N.}~\bibnamefont{Sandulescu}},
  \bibinfo{journal}{Nucl. Phys. A} \textbf{\bibinfo{volume}{789}},
  \bibinfo{pages}{94} (\bibinfo{year}{2007}).

\bibitem[{\citenamefont{Agrawal et~al.}(2000)\citenamefont{Agrawal, Sil, De,
  and Samaddar}}]{Agrawal2000}
\bibinfo{author}{\bibfnamefont{B.~K.} \bibnamefont{Agrawal}},
  \bibinfo{author}{\bibfnamefont{T.}~\bibnamefont{Sil}},
  \bibinfo{author}{\bibfnamefont{J.~N.} \bibnamefont{De}}, \bibnamefont{and}
  \bibinfo{author}{\bibfnamefont{S.~K.} \bibnamefont{Samaddar}},
  \bibinfo{journal}{Phys. Rev. C} \textbf{\bibinfo{volume}{62}},
  \bibinfo{pages}{044307} (\bibinfo{year}{2000}).

\bibitem[{\citenamefont{Lisboa et~al.}(2007)\citenamefont{Lisboa, Malhero, and
  Carlson}}]{Lisboa2007}
\bibinfo{author}{\bibfnamefont{R.}~\bibnamefont{Lisboa}},
  \bibinfo{author}{\bibfnamefont{M.}~\bibnamefont{Malhero}}, \bibnamefont{and}
  \bibinfo{author}{\bibfnamefont{B.~V.} \bibnamefont{Carlson}},
  \bibinfo{journal}{Int. J. Mod. Phys. E} \textbf{\bibinfo{volume}{16}},
  \bibinfo{pages}{3032} (\bibinfo{year}{2007}).

\bibitem[{\citenamefont{Lisboa et~al.}(2010)\citenamefont{Lisboa, Malheiro, and
  Carlson}}]{Lisboa2010}
\bibinfo{author}{\bibfnamefont{R.}~\bibnamefont{Lisboa}},
  \bibinfo{author}{\bibfnamefont{M.}~\bibnamefont{Malheiro}}, \bibnamefont{and}
  \bibinfo{author}{\bibfnamefont{B.~V.} \bibnamefont{Carlson}},
  \bibinfo{journal}{Nucl. Phys. B Proc. Suppl.} \textbf{\bibinfo{volume}{199}},
  \bibinfo{pages}{345} (\bibinfo{year}{2010}).

\bibitem[{\citenamefont{Kucharek and Ring}(1991)}]{Kucharek1991}
\bibinfo{author}{\bibfnamefont{H.}~\bibnamefont{Kucharek}} \bibnamefont{and}
  \bibinfo{author}{\bibfnamefont{P.}~\bibnamefont{Ring}}, \bibinfo{journal}{Z.
  Phys. A.} \textbf{\bibinfo{volume}{339}}, \bibinfo{pages}{23}
  (\bibinfo{year}{1991}).

\bibitem[{\citenamefont{Meng and Ring}(1996)}]{Meng1996}
\bibinfo{author}{\bibfnamefont{J.}~\bibnamefont{Meng}} \bibnamefont{and}
  \bibinfo{author}{\bibfnamefont{P.}~\bibnamefont{Ring}},
  \bibinfo{journal}{Phys. Rev. Lett.} \textbf{\bibinfo{volume}{77}},
  \bibinfo{pages}{3963} (\bibinfo{year}{1996}).

\bibitem[{\citenamefont{Gonzalez-Llarena
  et~al.}(1996)\citenamefont{Gonzalez-Llarena, Egido, Lalazissis, and
  Ring}}]{Gonzalez-Llarena1996}
\bibinfo{author}{\bibfnamefont{T.}~\bibnamefont{Gonzalez-Llarena}},
  \bibinfo{author}{\bibfnamefont{J.~L.} \bibnamefont{Egido}},
  \bibinfo{author}{\bibfnamefont{G.~A.} \bibnamefont{Lalazissis}},
  \bibnamefont{and} \bibinfo{author}{\bibfnamefont{P.}~\bibnamefont{Ring}},
  \bibinfo{journal}{Phys. Lett. B} \textbf{\bibinfo{volume}{379}},
  \bibinfo{pages}{13} (\bibinfo{year}{1996}).

\bibitem[{\citenamefont{Meng}(1998)}]{Meng1998NPA}
\bibinfo{author}{\bibfnamefont{J.}~\bibnamefont{Meng}}, \bibinfo{journal}{Nucl.
  Phys. A} \textbf{\bibinfo{volume}{635}}, \bibinfo{pages}{3}
  (\bibinfo{year}{1998}).

\bibitem[{\citenamefont{Serra and Ring}(2002)}]{Serra2002}
\bibinfo{author}{\bibfnamefont{M.}~\bibnamefont{Serra}} \bibnamefont{and}
  \bibinfo{author}{\bibfnamefont{P.}~\bibnamefont{Ring}},
  \bibinfo{journal}{Phys. Rev. C} \textbf{\bibinfo{volume}{65}},
  \bibinfo{pages}{064324} (\bibinfo{year}{2002}).

\bibitem[{\citenamefont{Nik\v{s}i\'{c}
  et~al.}(2002)\citenamefont{Nik\v{s}i\'{c}, Vretenar, Finelli, and
  Ring}}]{Niksic2002}
\bibinfo{author}{\bibfnamefont{T.}~\bibnamefont{Nik\v{s}i\'{c}}},
  \bibinfo{author}{\bibfnamefont{D.}~\bibnamefont{Vretenar}},
  \bibinfo{author}{\bibfnamefont{P.}~\bibnamefont{Finelli}}, \bibnamefont{and}
  \bibinfo{author}{\bibfnamefont{P.}~\bibnamefont{Ring}},
  \bibinfo{journal}{Phys. Rev. C} \textbf{\bibinfo{volume}{66}},
  \bibinfo{pages}{024306} (\bibinfo{year}{2002}).

\bibitem[{\citenamefont{Vretenar et~al.}(2005)\citenamefont{Vretenar,
  Afanasjev, Lalazissis, and Ring}}]{Vretenar2005}
\bibinfo{author}{\bibfnamefont{D.}~\bibnamefont{Vretenar}},
  \bibinfo{author}{\bibfnamefont{A.~V.} \bibnamefont{Afanasjev}},
  \bibinfo{author}{\bibfnamefont{G.~A.} \bibnamefont{Lalazissis}},
  \bibnamefont{and} \bibinfo{author}{\bibfnamefont{P.}~\bibnamefont{Ring}},
  \bibinfo{journal}{Phys. Rep.} \textbf{\bibinfo{volume}{409}},
  \bibinfo{pages}{101} (\bibinfo{year}{2005}).

\bibitem[{\citenamefont{Meng et~al.}(2006)\citenamefont{Meng, Toki, Zhou,
  Zhang, Long, and Geng}}]{Meng2006PPNP}
\bibinfo{author}{\bibfnamefont{J.}~\bibnamefont{Meng}},
  \bibinfo{author}{\bibfnamefont{H.}~\bibnamefont{Toki}},
  \bibinfo{author}{\bibfnamefont{S.~G.} \bibnamefont{Zhou}},
  \bibinfo{author}{\bibfnamefont{S.~Q.} \bibnamefont{Zhang}},
  \bibinfo{author}{\bibfnamefont{W.~H.} \bibnamefont{Long}}, \bibnamefont{and}
  \bibinfo{author}{\bibfnamefont{L.~S.} \bibnamefont{Geng}},
  \bibinfo{journal}{Prog. Part. Nucl. Phys.} \textbf{\bibinfo{volume}{57}},
  \bibinfo{pages}{470} (\bibinfo{year}{2006}).

\bibitem[{\citenamefont{Nik\v{s}i\'{c}
  et~al.}(2011)\citenamefont{Nik\v{s}i\'{c}, Vretenar, and Ring}}]{Niksic2011}
\bibinfo{author}{\bibfnamefont{T.}~\bibnamefont{Nik\v{s}i\'{c}}},
  \bibinfo{author}{\bibfnamefont{D.}~\bibnamefont{Vretenar}}, \bibnamefont{and}
  \bibinfo{author}{\bibfnamefont{P.}~\bibnamefont{Ring}},
  \bibinfo{journal}{Prog. Part. Nucl. Phys.} \textbf{\bibinfo{volume}{66}},
  \bibinfo{pages}{519 } (\bibinfo{year}{2011}).

\bibitem[{\citenamefont{Zhou et~al.}(2010)\citenamefont{Zhou, Meng, Ring, and
  Zhao}}]{Zhou2010}
\bibinfo{author}{\bibfnamefont{S.-G.} \bibnamefont{Zhou}},
  \bibinfo{author}{\bibfnamefont{J.}~\bibnamefont{Meng}},
  \bibinfo{author}{\bibfnamefont{P.}~\bibnamefont{Ring}}, \bibnamefont{and}
  \bibinfo{author}{\bibfnamefont{E.-G.} \bibnamefont{Zhao}},
  \bibinfo{journal}{Phys. Rev. C} \textbf{\bibinfo{volume}{82}},
  \bibinfo{pages}{011301(R)} (\bibinfo{year}{2010}).

\bibitem[{\citenamefont{Li et~al.}(2012{\natexlab{a}})\citenamefont{Li, Meng,
  Ring, Zhao, and Zhou}}]{Li2012}
\bibinfo{author}{\bibfnamefont{L.}~\bibnamefont{Li}},
  \bibinfo{author}{\bibfnamefont{J.}~\bibnamefont{Meng}},
  \bibinfo{author}{\bibfnamefont{P.}~\bibnamefont{Ring}},
  \bibinfo{author}{\bibfnamefont{E.-G.} \bibnamefont{Zhao}}, \bibnamefont{and}
  \bibinfo{author}{\bibfnamefont{S.-G.} \bibnamefont{Zhou}},
  \bibinfo{journal}{Phys. Rev. C} \textbf{\bibinfo{volume}{85}},
  \bibinfo{pages}{024312} (\bibinfo{year}{2012}{\natexlab{a}}).

\bibitem[{\citenamefont{Li et~al.}(2012{\natexlab{b}})\citenamefont{Li, Meng,
  Ring, Zhao, and Zhou}}]{Li2012CPL}
\bibinfo{author}{\bibfnamefont{L.-L.} \bibnamefont{Li}},
  \bibinfo{author}{\bibfnamefont{J.}~\bibnamefont{Meng}},
  \bibinfo{author}{\bibfnamefont{P.}~\bibnamefont{Ring}},
  \bibinfo{author}{\bibfnamefont{E.-G.} \bibnamefont{Zhao}}, \bibnamefont{and}
  \bibinfo{author}{\bibfnamefont{S.-G.} \bibnamefont{Zhou}},
  \bibinfo{journal}{Chin. Phys. Lett.} \textbf{\bibinfo{volume}{29}},
  \bibinfo{pages}{042101} (\bibinfo{year}{2012}{\natexlab{b}}).

\bibitem[{\citenamefont{Chen et~al.}(2012)\citenamefont{Chen, Li, Liang, and
  Meng}}]{Chen2012}
\bibinfo{author}{\bibfnamefont{Y.}~\bibnamefont{Chen}},
  \bibinfo{author}{\bibfnamefont{L.}~\bibnamefont{Li}},
  \bibinfo{author}{\bibfnamefont{H.}~\bibnamefont{Liang}}, \bibnamefont{and}
  \bibinfo{author}{\bibfnamefont{J.}~\bibnamefont{Meng}},
  \bibinfo{journal}{Phys. Rev. C} \textbf{\bibinfo{volume}{85}},
  \bibinfo{pages}{067301} (\bibinfo{year}{2012}).

\bibitem[{\citenamefont{P\"oschl et~al.}(1997)\citenamefont{P\"oschl, Vretenar,
  Lalazissis, and Ring}}]{Poschl1997}
\bibinfo{author}{\bibfnamefont{W.}~\bibnamefont{P\"oschl}},
  \bibinfo{author}{\bibfnamefont{D.}~\bibnamefont{Vretenar}},
  \bibinfo{author}{\bibfnamefont{G.~A.} \bibnamefont{Lalazissis}},
  \bibnamefont{and} \bibinfo{author}{\bibfnamefont{P.}~\bibnamefont{Ring}},
  \bibinfo{journal}{Phys. Rev. Lett.} \textbf{\bibinfo{volume}{79}},
  \bibinfo{pages}{3841} (\bibinfo{year}{1997}).

\bibitem[{\citenamefont{Meng and Ring}(1998)}]{Meng1998}
\bibinfo{author}{\bibfnamefont{J.}~\bibnamefont{Meng}} \bibnamefont{and}
  \bibinfo{author}{\bibfnamefont{P.}~\bibnamefont{Ring}},
  \bibinfo{journal}{Phys. Rev. Lett.} \textbf{\bibinfo{volume}{80}},
  \bibinfo{pages}{460} (\bibinfo{year}{1998}).


\bibitem[{\citenamefont{Meng et~al.}(2002)\citenamefont{Meng,  Toki, Zeng, Zhang, and Zhou}}]{Meng2002PRC}
\bibinfo{author}{\bibfnamefont{J.}~\bibnamefont{Meng}},
  \bibinfo{author}{\bibfnamefont{H.}~\bibnamefont{Toki}},
  \bibinfo{author}{\bibfnamefont{J.~Y.}~\bibnamefont{Zeng}},
  \bibinfo{author}{\bibfnamefont{S.~Q.}~\bibnamefont{Zhang}}, \bibnamefont{and}
  \bibinfo{author}{\bibfnamefont{S.~-G.}~\bibnamefont{Zhou}},
  \bibinfo{journal}{Phys. Rev. C} \textbf{\bibinfo{volume}{65}},
  \bibinfo{pages}{041302} (\bibinfo{year}{2002}).




\bibitem[{\citenamefont{Vretenar et~al.}(1999)\citenamefont{Vretenar,
  Lalazissis, and Ring}}]{vretenar1999}
\bibinfo{author}{\bibfnamefont{D.}~\bibnamefont{Vretenar}},
  \bibinfo{author}{\bibfnamefont{G.~A.} \bibnamefont{Lalazissis}},
  \bibnamefont{and} \bibinfo{author}{\bibfnamefont{P.}~\bibnamefont{Ring}},
  \bibinfo{journal}{Phys. Rev. Lett.} \textbf{\bibinfo{volume}{82}},
  \bibinfo{pages}{4595} (\bibinfo{year}{1999}).

\bibitem[{\citenamefont{Lalazissis et~al.}(1998)\citenamefont{Lalazissis,
  Vretenar, W., and Ring}}]{Lalazissis1998}
\bibinfo{author}{\bibfnamefont{G.}~\bibnamefont{Lalazissis}},
  \bibinfo{author}{\bibfnamefont{D.}~\bibnamefont{Vretenar}},
  \bibinfo{author}{\bibfnamefont{P.}~\bibnamefont{W.}}, \bibnamefont{and}
  \bibinfo{author}{\bibfnamefont{P.}~\bibnamefont{Ring}},
  \bibinfo{journal}{Phys. Lett. B} \textbf{\bibinfo{volume}{418}},
  \bibinfo{pages}{7} (\bibinfo{year}{1998}).

\bibitem[{\citenamefont{Lalazissis et~al.}(1999)\citenamefont{Lalazissis,
  Vretenar, Ring, Stoitsov, and Robledo}}]{Lalazissis1999}
\bibinfo{author}{\bibfnamefont{G.~A.} \bibnamefont{Lalazissis}},
  \bibinfo{author}{\bibfnamefont{D.}~\bibnamefont{Vretenar}},
  \bibinfo{author}{\bibfnamefont{P.}~\bibnamefont{Ring}},
  \bibinfo{author}{\bibfnamefont{M.}~\bibnamefont{Stoitsov}}, \bibnamefont{and}
  \bibinfo{author}{\bibfnamefont{L.~M.} \bibnamefont{Robledo}},
  \bibinfo{journal}{Phys. Rev. C} \textbf{\bibinfo{volume}{60}},
  \bibinfo{pages}{014310} (\bibinfo{year}{1999}).

\bibitem[{\citenamefont{Meng et~al.}(1998)\citenamefont{Meng, Sugawara-Tanabe,
  Yamaji, Ring, and Arima}}]{Meng1998PRC}
\bibinfo{author}{\bibfnamefont{J.}~\bibnamefont{Meng}},
  \bibinfo{author}{\bibfnamefont{K.}~\bibnamefont{Sugawara-Tanabe}},
  \bibinfo{author}{\bibfnamefont{S.}~\bibnamefont{Yamaji}},
  \bibinfo{author}{\bibfnamefont{P.}~\bibnamefont{Ring}}, \bibnamefont{and}
  \bibinfo{author}{\bibfnamefont{A.}~\bibnamefont{Arima}},
  \bibinfo{journal}{Phys. Rev. C} \textbf{\bibinfo{volume}{58}},
  \bibinfo{pages}{R628} (\bibinfo{year}{1998}).

\bibitem[{\citenamefont{Meng et~al.}(1999)\citenamefont{Meng, Sugawara-Tanabe,
  Yamaji, and Arima}}]{Meng1999}
\bibinfo{author}{\bibfnamefont{J.}~\bibnamefont{Meng}},
  \bibinfo{author}{\bibfnamefont{K.}~\bibnamefont{Sugawara-Tanabe}},
  \bibinfo{author}{\bibfnamefont{S.}~\bibnamefont{Yamaji}}, \bibnamefont{and}
  \bibinfo{author}{\bibfnamefont{A.}~\bibnamefont{Arima}},
  \bibinfo{journal}{Phys. Rev. C} \textbf{\bibinfo{volume}{59}},
  \bibinfo{pages}{154} (\bibinfo{year}{1999}).

\bibitem[{\citenamefont{Zhang et~al.}(2005)\citenamefont{Zhang, Meng, Zhang,
  Geng, and Toki}}]{Zhang2005}
\bibinfo{author}{\bibfnamefont{W.}~\bibnamefont{Zhang}},
  \bibinfo{author}{\bibfnamefont{J.}~\bibnamefont{Meng}},
  \bibinfo{author}{\bibfnamefont{S.~Q.} \bibnamefont{Zhang}},
  \bibinfo{author}{\bibfnamefont{L.~S.} \bibnamefont{Geng}}, \bibnamefont{and}
  \bibinfo{author}{\bibfnamefont{H.}~\bibnamefont{Toki}},
  \bibinfo{journal}{Nucl. Phys. A} \textbf{\bibinfo{volume}{753}},
  \bibinfo{pages}{106} (\bibinfo{year}{2005}).

\bibitem[{\citenamefont{Nikolaus, Hoch, and Madland}(1992)}]{Nikolaus1992}
\bibinfo{author}{\bibfnamefont{B.~A.} \bibnamefont{Nikolaus}},
  \bibinfo{author}{\bibfnamefont{T.}~\bibnamefont{Hoch}}, \bibnamefont{and}
  \bibinfo{author}{\bibfnamefont{D.~G.} \bibnamefont{Madland}},
  \bibinfo{journal}{Phys. Rev. C} \textbf{\bibinfo{volume}{46}},
  \bibinfo{pages}{1757} (\bibinfo{year}{1992}).

\bibitem[{\citenamefont{B\"urvenich et~al.}(2002)\citenamefont{B\"urvenich,
  Madland, Maruhn, and Reinhard}}]{Burvenich2002}
\bibinfo{author}{\bibfnamefont{T.}~\bibnamefont{B\"urvenich}},
  \bibinfo{author}{\bibfnamefont{D.~G.} \bibnamefont{Madland}},
  \bibinfo{author}{\bibfnamefont{J.~A.} \bibnamefont{Maruhn}},
  \bibnamefont{and} \bibinfo{author}{\bibfnamefont{P.-G.}
  \bibnamefont{Reinhard}}, \bibinfo{journal}{Phys. Rev. C}
  \textbf{\bibinfo{volume}{65}}, \bibinfo{pages}{044308}
  (\bibinfo{year}{2002}).

\bibitem[{\citenamefont{Nik\v{s}i\'{c}
  et~al.}(2008)\citenamefont{Nik\v{s}i\'{c}, Vretenar, and Ring}}]{Niksic2008}
\bibinfo{author}{\bibfnamefont{T.}~\bibnamefont{Nik\v{s}i\'{c}}},
  \bibinfo{author}{\bibfnamefont{D.}~\bibnamefont{Vretenar}}, \bibnamefont{and}
  \bibinfo{author}{\bibfnamefont{P.}~\bibnamefont{Ring}},
  \bibinfo{journal}{Phys. Rev. C} \textbf{\bibinfo{volume}{78}},
  \bibinfo{pages}{034318} (\bibinfo{year}{2008}).

\bibitem[{\citenamefont{Zhao et~al.}(2010)\citenamefont{Zhao, Li, Yao, and
  Meng}}]{Zhao2010}
\bibinfo{author}{\bibfnamefont{P.~W.} \bibnamefont{Zhao}},
  \bibinfo{author}{\bibfnamefont{Z.~P.} \bibnamefont{Li}},
  \bibinfo{author}{\bibfnamefont{J.~M.} \bibnamefont{Yao}}, \bibnamefont{and}
  \bibinfo{author}{\bibfnamefont{J.}~\bibnamefont{Meng}},
  \bibinfo{journal}{Phys. Rev. C} \textbf{\bibinfo{volume}{82}},
  \bibinfo{pages}{054319} (\bibinfo{year}{2010}).

\bibitem[{\citenamefont{Mei et~al.}(2012)\citenamefont{Mei, Xiang, Yao, Li, and
  Meng}}]{Mei2012}
\bibinfo{author}{\bibfnamefont{H.}~\bibnamefont{Mei}},
  \bibinfo{author}{\bibfnamefont{J.}~\bibnamefont{Xiang}},
  \bibinfo{author}{\bibfnamefont{J.~M.} \bibnamefont{Yao}},
  \bibinfo{author}{\bibfnamefont{Z.~P.} \bibnamefont{Li}}, \bibnamefont{and}
  \bibinfo{author}{\bibfnamefont{J.}~\bibnamefont{Meng}},
  \bibinfo{journal}{Phys. Rev. C} \textbf{\bibinfo{volume}{85}},
  \bibinfo{pages}{034321} (\bibinfo{year}{2012}).

\bibitem[{\citenamefont{Hua et~al.}(2012)\citenamefont{Hua, Heng, Niu, Sun, and
  Guo}}]{Hua2012}
\bibinfo{author}{\bibfnamefont{X.~M.}~\bibnamefont{Hua}},
  \bibinfo{author}{\bibfnamefont{T.~H.} \bibnamefont{Heng}},
  \bibinfo{author}{\bibfnamefont{Z.~M.} \bibnamefont{Niu}},
  \bibinfo{author}{\bibfnamefont{B.~H.} \bibnamefont{Sun}}, \bibnamefont{and}
  \bibinfo{author}{\bibfnamefont{J.~Y.} \bibnamefont{Guo}},
  \bibinfo{journal}{Sci. China Phys. Mech. Astron.}
  \textbf{\bibinfo{volume}{55}}, \bibinfo{pages}{2414} (\bibinfo{year}{2012}).

\bibitem[{\citenamefont{Niu et~al.}(2013)\citenamefont{Niu, Niu, Liu, Liang,
  and Guo}}]{Niu2013}
\bibinfo{author}{\bibfnamefont{Z.~M.} \bibnamefont{Niu}},
  \bibinfo{author}{\bibfnamefont{Y.~F.} \bibnamefont{Niu}},
  \bibinfo{author}{\bibfnamefont{Q.}~\bibnamefont{Liu}},
  \bibinfo{author}{\bibfnamefont{H.~Z.} \bibnamefont{Liang}}, \bibnamefont{and}
  \bibinfo{author}{\bibfnamefont{J.~Y.} \bibnamefont{Guo}},
  \bibinfo{journal}{Phys. Rev. C} \textbf{\bibinfo{volume}{87}},
  \bibinfo{pages}{051303(R)} (\bibinfo{year}{2013}).

\bibitem[{\citenamefont{Decharg\'e and Gogny}(1980)}]{Decharge1980}
\bibinfo{author}{\bibfnamefont{J.}~\bibnamefont{Decharg\'e}} \bibnamefont{and}
  \bibinfo{author}{\bibfnamefont{D.}~\bibnamefont{Gogny}},
  \bibinfo{journal}{Phys. Rev. C} \textbf{\bibinfo{volume}{21}},
  \bibinfo{pages}{1568} (\bibinfo{year}{1980}).

\bibitem[{\citenamefont{Berger et~al.}(1984)\citenamefont{Berger, Girod, and
  Gogny}}]{Berger1984}
\bibinfo{author}{\bibfnamefont{J.}~\bibnamefont{Berger}},
  \bibinfo{author}{\bibfnamefont{M.}~\bibnamefont{Girod}}, \bibnamefont{and}
  \bibinfo{author}{\bibfnamefont{D.}~\bibnamefont{Gogny}},
  \bibinfo{journal}{Nucl. Phys. A} \textbf{\bibinfo{volume}{428}},
  \bibinfo{pages}{23 } (\bibinfo{year}{1984}).

\bibitem[{\citenamefont{Berger et~al.}(1991)\citenamefont{Berger, Girod, and
  Gogny}}]{Berger1991}
\bibinfo{author}{\bibfnamefont{J.~F.} \bibnamefont{Berger}},
  \bibinfo{author}{\bibfnamefont{M.}~\bibnamefont{Girod}}, \bibnamefont{and}
  \bibinfo{author}{\bibfnamefont{D.}~\bibnamefont{Gogny}},
  \bibinfo{journal}{Comput. Phys. Commun.} \textbf{\bibinfo{volume}{63}},
  \bibinfo{pages}{365} (\bibinfo{year}{1991}).

\bibitem[{\citenamefont{Kucharek et~al.}(1989)\citenamefont{Kucharek, Ring,
  Schuck, Bengtsson, and Girod}}]{Kucharek1989}
\bibinfo{author}{\bibfnamefont{H.}~\bibnamefont{Kucharek}},
  \bibinfo{author}{\bibfnamefont{P.}~\bibnamefont{Ring}},
  \bibinfo{author}{\bibfnamefont{P.}~\bibnamefont{Schuck}},
  \bibinfo{author}{\bibfnamefont{R.}~\bibnamefont{Bengtsson}},
  \bibnamefont{and} \bibinfo{author}{\bibfnamefont{M.}~\bibnamefont{Girod}},
  \bibinfo{journal}{Phys. Lett. B} \textbf{\bibinfo{volume}{216}},
  \bibinfo{pages}{249} (\bibinfo{year}{1989}).

\bibitem[{\citenamefont{Serra et~al.}(2001)\citenamefont{Serra, Rummel, and
  Ring}}]{Serra2001}
\bibinfo{author}{\bibfnamefont{M.}~\bibnamefont{Serra}},
  \bibinfo{author}{\bibfnamefont{A.}~\bibnamefont{Rummel}}, \bibnamefont{and}
  \bibinfo{author}{\bibfnamefont{P.}~\bibnamefont{Ring}},
  \bibinfo{journal}{Phys. Rev. C} \textbf{\bibinfo{volume}{65}},
  \bibinfo{pages}{014304} (\bibinfo{year}{2001}).

\bibitem[{\citenamefont{Tian et~al.}(2009)\citenamefont{Tian, Ma, and
  Ring}}]{Tian2009}
\bibinfo{author}{\bibfnamefont{Y.}~\bibnamefont{Tian}},
  \bibinfo{author}{\bibfnamefont{Z.~Y.} \bibnamefont{Ma}}, \bibnamefont{and}
  \bibinfo{author}{\bibfnamefont{P.}~\bibnamefont{Ring}},
  \bibinfo{journal}{Phys. Lett. B} \textbf{\bibinfo{volume}{676}},
  \bibinfo{pages}{44} (\bibinfo{year}{2009}).

\bibitem[{\citenamefont{Bohr and Mottelson}(1969)}]{BohrMottelsonbook}
\bibinfo{author}{\bibfnamefont{A.}~\bibnamefont{Bohr}} \bibnamefont{and}
  \bibinfo{author}{\bibfnamefont{B.~R.} \bibnamefont{Mottelson}},
  \emph{\bibinfo{title}{Nuclear Structure}}, vol.~\bibinfo{volume}{I}
  (\bibinfo{publisher}{Benjamin Inc., New York}, \bibinfo{year}{1969}).

\bibitem[{\citenamefont{Bethe}(1936)}]{Bethe1936}
\bibinfo{author}{\bibfnamefont{H.~A.} \bibnamefont{Bethe}},
  \bibinfo{journal}{Phys. Rev.} \textbf{\bibinfo{volume}{50}},
  \bibinfo{pages}{332} (\bibinfo{year}{1936}).

\bibitem[{\citenamefont{Bonche et~al.}(1984)\citenamefont{Bonche, Levit, and
  Vautherin}}]{Bonche1984}
\bibinfo{author}{\bibfnamefont{P.}~\bibnamefont{Bonche}},
  \bibinfo{author}{\bibfnamefont{S.}~\bibnamefont{Levit}}, \bibnamefont{and}
  \bibinfo{author}{\bibfnamefont{D.}~\bibnamefont{Vautherin}},
  \bibinfo{journal}{Nucl. Phys. A} \textbf{\bibinfo{volume}{428}},
  \bibinfo{pages}{95} (\bibinfo{year}{1984}).

\end{thebibliography}
%

\end{document}